\documentclass[12pt]{article}

\usepackage{epsfig}

\usepackage{amssymb}
\usepackage{amsmath}
\usepackage{amsfonts}

\def\be{\begin{equation}}
\def\ee{\end{equation}}
\def\ba{\begin{array}{c}}
\def\ea{\end{array}}

\def\ben{$$}
\def\een{$$}

\newcommand{\bea}{\begin{eqnarray}}
\newcommand{\eea}{\end{eqnarray}}

\newcommand{\bbr}{\br\!\br}

\newcommand{\kt}{\rangle}
\newcommand{\br}{\langle}
\begin{document}

\titlepage

\vspace{.35cm}

 \begin{center}{\Large \bf
Fundamental length in quantum theories with ${\cal PT}-$symmetric
Hamiltonians II:

The case of quantum graphs.


  }\end{center}

\vspace{10mm}

 \begin{center}

 {\bf Miloslav Znojil}

 \vspace{3mm}
Nuclear Physics Institute ASCR,

250 68 \v{R}e\v{z}, Czech Republic

{e-mail: znojil@ujf.cas.cz}

\vspace{3mm}

\end{center}

\vspace{5mm}

\section*{Abstract}

${\cal PT}-$symmetrization of quantum graphs is proposed as an
innovation where an adjustable, tunable nonlocality is admitted. The
proposal generalizes the ${\cal PT}-$symmetric square-well models of
Ref.~\cite{fund} (with real spectrum and with a variable fundamental
length $\theta$) which are reclassified as the most elementary
quantum $q-$pointed-star graphs with minimal $q=2$. Their
equilateral $q=3,4,\ldots$ generalizations are considered, with
interactions attached to the vertices. Runge-Kutta discretization of
coordinates simplifies the quantitative analysis by reducing our
graphs to star-shaped lattices of $N=qK+1$ points. The resulting
bound-state spectra are found real in an $N-$independent interval of
couplings ${\lambda}\in (-1,1)$. Inside this interval the set of
closed-form metrics $\Theta^{(N)}_j({\lambda})$ is constructed,
defining independent eligible local (at $j=0$) or increasingly
nonlocal (at $j=1,2, \ldots$) inner products in the respective
physical Hilbert spaces of states ${\cal H}^{(N)}_j({\lambda})$. In
this way each graph is assigned a menu of non-equivalent, optional
probabilistic quantum interpretations.

\newpage

\section{Introduction}

Many nontrivial quantum systems are described via a simplified
effective model. Vibrational excitations of fields, nuclei or
molecules may often be represented, for example, by artificial
models where a single real or virtual (quasi)particle moves along
a suitable one-dimensional trajectory, finite or infinite. In
paper I \cite{fund} we even analyzed a family of models where this
trajectory has further been replaced, in the so called Runge-Kutta
approximation, by a finite lattice of points.

Specific difficulties may survive even after a drastic reduction
of the number of degrees of freedom. Typically, a
quasi-one-dimensional narrow-tube trajectory may happen to be
curved (causing the emergence of bound states \cite{Seba}) or
twisted (returning these bound states to the free-motion continuum
again \cite{Krejcirik}). Other pathologies may emerge when the
(quasi)particle moves along a topologically nontrivial waveguide.
In the simplest thin-tube realization of the latter scenario one
speaks, in general, about the motion of a quantum (quasi)particle
along a {\em graph}, i.e., along a system of one-dimensional
free-motion trajectories (called ``edges" of the graph) connected
at the so called ``vertices" of the graph (various nontrivial
interactions could be admitted at these points).

A purely phenomenological motivation of interest in quantum graphs
has originally emerged in quantum chemistry where the edges were
identified with the bonds between atoms in a larger organic
molecule along which the electrons might move almost freely
\cite{napohtal}. Soon, a more abstract appeal of quantum graphs
prevailed offering a nontrivial quantitative picture of quantum
dynamics in many arrangements ranging from the  Y-shaped tree up
to fractal trajectories \cite{fractals}. The recent proceedings
\cite{exproc} can be cited as a source of updated information
about the current state of the art. More than 700 pages of
predominantly mathematically oriented reviews still incorporate a
few physics-centered summaries of potentially appealing
phenomenological consequences and applications of the theory. {\it
Pars pro toto} we could point out refs.~\cite{Kuchment,Exner})
putting more emphasis on physics and listing many related
references. Today, the use of quantum graphs ranges from the
analysis of photonic crystals up to the studies of thin wires and
waveguides and other mesoscopic devices produced by sophisticated
nanotechnologies. On theoretical level quantum graphs are
increasingly popular as formal structures testing field theory
\cite{Ragou}  or describing certain important phenomena in
solid-state physics \cite{Hatano3}. Multiple concrete models serve
as a laboratory of our understanding of systems with constraints
\cite{Brody}. Last but not least one finds quantum graphs used as
benchmark systems in quantum chaos \cite{Smil} and/or random walks
\cite{Huill}.

The incessant transfer of the quantum-graph idea from its original,
purely descriptive role to a more abstract theoretical framework may
be expected to continue. An illustration of the emergence of new
tendencies in this field may be seen, e.g., in the {\em
complexified}, non-Hermitian  boundary-supported interactions as
studied, in the context of fully realistic three-dimensional lattice
models, in Ref.~\cite{Molinari}. These tendencies grew from origins
which may be traced back to a few papers by Hatano and Nelson
\cite{Hatano} and by Feinberg and Zee \cite{Hatano2} as well as to a
number of more recent studies rooted not only in solid-state physics
\cite{Hatano3} but also, say, in nuclear physics \cite{Geyer}, field
theory \cite{Kleefeld} or cosmology \cite{Andrianov}. An
intensification of interest in all of these non-Hermitian quantum
systems with real spectra occurred, in particular, after the
publication of influential letter \cite{BB} where the very special
form of non-Hermiticity called ${\cal PT}-$symmetry of Hamiltonians
(which is to be explained below) has been promoted as an
unexpectedly productive heuristic principle. It found many concrete
applications reported, e.g., in proceedings of several dedicated
conferences \cite{proc}. Virtually all of these studies may be
characterized as a search for a new point of optimal balance between
the mathematical requirements of simplicity (and, in particular, of
constructive tractability of physical models) and the natural
requirements of dynamical and phenomenological relevance of new
models. For illustration let us mention just the recent proposals of
the tests of quantum brachistochrones (where the recent letter
\cite{Uwe} summarizes the existing theoretical proposals) or of
measurements over certain anomalous scattering systems
\cite{alispectralsing}. The quickly developing discussion of
possible measurable effects involving ${\cal PT}-$symmetric systems
in quantum optics \cite{optics} (and, perhaps, in quantum gravity
etc \cite{Bijan}) must also certainly be mentioned here.

The emergence of all of these new theoretical ideas motivated also
our present work. Their multisided applicability persuaded us not
only about an undeniable phenomenological appeal and relevance of
non-Hermitian interaction models but also about the promising
tractability and feasibility of many of their computational and
constructive aspects. In what follows we shall propose and study,
therefore, a schematic though still nontrivial quantum-graph models
based on a non-Hermitian form of interaction supported, as usual,
just by certain vertices of the given graph.

The text will start from a concise outline of the inspiration and
origins incorporating simple square-well models reviewed in
Section \ref{secII} and their elementary Hermitian star-shaped
discrete quantum graph generalizations proposed in Section
\ref{secIII}. The key ideas of our present innovations will be
then listed in Section \ref{secIV} followed by Section \ref{reali}
where the main necessary property of our quantum graph models,
viz., the reality of their spectra will be demonstrated. Section
\ref{blokybe} will then be devoted to the presentation and
explanation of the core of our message, viz., to the description
of a few first nontrivial examples of nonlocal ${\cal
PT}-$symmetric quantum graphs. Finally, our concluding remarks
will be collected in Section \ref{finis}.


\section{Square-well Schr\"{o}dinger equations \label{secII}}

\subsection{Runge-Kutta discretization }

Let us start our considerations from the most common ordinary
differential Schr\"{o}dinger equation for bound states in a square
well,
 \be
 -\frac{d^2}{d\xi^2}\,\psi(\xi) =E\,\psi(\xi)\,,
 \ \ \ \ \ \ \ \psi(\pm L)
  =0\,
 \label{SEsqw}
 \ee
and  review a few results obtained for various modifications and
perturbations of this model in the recent literature. First, let us
mention the study \cite{Tretter} where the addition of a
``sufficiently small" potential $V(\xi)$ has been shown to leave the
spectrum real (i.e., in principle, observable) {\em irrespectively}
of the detailed form of function $V(\xi)$ which can even be allowed
{\em complex}. This result can be perceived as one of the most
persuasive rigorous mathematical confirmations of the Bender's and
Boettcher's conjecture \cite{BB} that in applied Quantum Mechanics
there exists a broad class of complex potentials supporting {\em
real} spectra of bound-state energies.

Bender et al  \cite{BBjmp}  noticed and emphasized that many
Hamiltonians $H=p^2+V(\xi)\neq H^\dagger$ can be characterized by
their ${\cal PT}-$symmetry, i.e., by the property $H{\cal PT}={\cal
PT}H$ with parity ${\cal P}$ and with time reversal ${\cal T}$
mimicked by Hermitian conjugation \cite{erratum}. Bound states in a
few solvable ${\cal PT}-$symmetric piecewise-constant potentials
were studied in Refs.~\cite{ptsqw}. The correct probabilistic
interpretation of some of these potentials found its first
constructive formulation in Ref.~\cite{Batal}. In parallel, an
efficient simplification of the underlying mathematics via
Runge-Kutta (RK) discretization of coordinates has been proposed in
Refs.~\cite{Weigert}. It was based on the replacement of the
interval of $\xi \in (-L,L)$ by its discrete version
 \be
  \xi_k=k\,h\,,
 \ \ \ \
 \ \ \ \ k = 0, \pm 1, \ldots,
 \pm K\,,\ \ \ \ h>0\,
 \label{RKL}
 \ee
i.e., by the discrete lattice of points
 \be
 \ba
  \begin{array}{|c|}
 \hline
 {\xi_{-K}}\\
 \hline
 \ea \ {\xi_{-K+1}}\
 \ldots\   \xi_{-2}\  \xi_{-1}\
 \begin{array}{||c||}
 \hline
 \hline
 {\xi_0}\\
 \hline
 \hline
 \ea
 \  \xi_1\  \xi_2\  \ldots\ {\xi_{K-1}}\
   \begin{array}{|c|}
 \hline
 {\xi_{K}}\\
 \hline
 \ea  \\
  \ea\,.
 \label{VgstRKL}
 \ee
In this perspective one has to replace differential
Eq.~(\ref{SEsqw}) by its discrete analogue or approximation
 \be
 -\frac{\psi(\xi_{k-1})-2\,\psi(\xi_k)+\psi(\xi_{k+1})}{h^2}
 =E\,\psi(\xi_k)\,.
 \label{SEdis}
 \ee
A clear insight in the formal structure of the square-well
eigenvalue problem is  achieved. With $-K \leq k \leq K$ and
$\psi(\xi_{\pm (K+1)})=0$ our difference Schr\"{o}dinger
Eq.~(\ref{SEdis}) reads
 \be
  \left[ \begin {array}{ccccccc}
   2&-1&0&\ldots&&\ldots&0
  \\
  -1
&2&-1&0&\ldots&\ldots&0
 \\0&-1&2&-1&\ddots&&\vdots
 \\
 \vdots&\ddots&\ddots&\ddots&\ddots&0&0
 \\
 {}&&
&-1&2&-1&0
 \\{}\vdots&&&\ddots&-1&2&-1
 \\{}0&\ldots&&\ldots&0&-1&2\\
 \end {array} \right]\,
 \left[ \begin {array}{c}
 \psi(\xi_{-K})\\
 \psi(\xi_{-K+1})\\
 \psi(\xi_{-K+2})\\
 \vdots\\
 \psi(\xi_{K-1})\\
 \psi(\xi_{K})\\
 \ea
 \right ]=E\,
 \left[ \begin {array}{c}
 \psi(\xi_{-K})\\
 \psi(\xi_{-K+1})\\
 \psi(\xi_{-K+2})\\
 \vdots\\
 \psi(\xi_{K-1})\\
 \psi(\xi_{K})\\
 \ea
 \right ]\,
 \label{kinetie}
 \ee
i.e., it acquires the transparent matrix-diagonalization form.

\subsection{Equivalence to a linear discrete quantum graph}

Let us renumber the linear array (\ref{VgstRKL}) of $N=2K+1$ points
in a slightly unusual manner which emphasizes its left-right
symmetry,
 \be
 \ba
  \begin{array}{|c|}
 \hline
 {x_{2K-1}}\\
 \hline
 \ea \ {x_{2K-3}}\
 \ldots\   x_3\  x_1\
 \begin{array}{||c||}
 \hline
 \hline
 {x_0}\\
 \hline
 \hline
 \ea
 \  x_2\  x_4\  \ldots\ {x_{2K-2}}\
   \begin{array}{|c|}
 \hline
 {x_{2K}}\\
 \hline
 \ea  \\
  \ea\,.
 \label{VgRKL}
 \ee
Our Schr\"{o}dinger Eq.~(\ref{kinetie}) becomes rearranged,
 \be
  \left[ \begin {array}{c|cccccc}
   2&-1&
 \multicolumn{1}{c}{-1}&0&0&\ldots&0
  \\
  \hline
  -1
&2&\multicolumn{1}{c}{0}&-1&0&\ddots&\vdots
 \\-1&0&\multicolumn{1}{c}{2}&0&\ddots&\ddots&\vdots
 \\
 0&-1&0&\ddots&\ddots&-1&0
 \\
  \vdots&\ddots&\ddots
&\ddots&2&0&-1
\\
 0&\ldots&\ddots&-1&0&2&0
 \\
 0&\ldots&\ldots&0&-1&0&2\\
 \end {array} \right]\,
 \left[ \begin {array}{c}
 \psi(x_0)\\
 \psi(x_1)\\
 \psi(x_2)\\
 \vdots\\
 \psi(x_{2K-1})\\
 \psi(x_{2K})\\
 \ea
 \right ]=E\,
 \left[ \begin {array}{c}
 \psi(x_0)\\
 \psi(x_1)\\
 \psi(x_2)\\
 \vdots\\
 \psi(x_{2K-1})\\
 \psi(x_{2K})\\
 \ea
 \right ]\,.
 \label{kineV}
 \ee
It may be perceived as describing a system which lives on the linear
(one could also call it V-shaped) graph which consists of two wedges
connected in the origin.

The $N-$dimensional Hamiltonian with $N=2K+1$ as it appears in
Eq.~(\ref{kineV}) has a block-tridiagonal partitioned matrix
structure
  \be
 H^{(N)}=
 \left[
  \begin {array}{c|ccccc}
 u&\vec{v}&\vec{0}&\cdots&\cdots&\vec{0}
 \\
 \hline
 \vec{v}^T&2I&-I&0&\ldots&0
 \\
 \vec{0}^T&-I&\ddots&\ddots&\ddots&\vdots
 \\
 \vec{0}^T&0&\ddots&2I&-I&0
 \\
 \vdots&\vdots&\ddots&-I&2I&-I
 \\
 \vec{0}^T&0&\cdots&0&-I&2I
 \end {array} \right]\,
 \label{blqpkk}
 \ee
with $\vec{v}=(-1,-1)$ and $\vec{0}=(0,0)$ being two-dimensional row
vectors while $u=2$ is a number. The rest of the matrix is composed
of two-dimensional unit matrices $I$ and null-matrices $0$. In the
light of what has been written in Introduction the naive, discrete
quantum square-well problem may be reinterpreted as one of the
simplest quantum graphs, therefore.

\section{Star-shaped discrete quantum graphs \label{secIII}}

The example of preceding section may be complemented by a series of
its generalizations living on $q-$pointed star graphs with
$q=3,4,\ldots$. In this new context the trivial example
(\ref{VgRKL}) + (\ref{kineV}) indicates how this generalization can
be ``translated" back into the language of difference or matrix
Schr\"{o}dinger equations. Let us now complement this idea by a few
concrete examples of its implementation.

\subsection{Y-shaped model: $q=3$}

The simplest nontrivial discrete realization of a graph with $q=3$
may be visualized as  an Y-shaped (or, if you wish, T-shaped)
$N-$point lattice composed of three equally long branches whose
individual points will be numbered as follows,
 \be
 \ba
  \begin{array}{|c|}
 \hline
 {x_{N-2}}\\
 \hline
 \ea\ {x_{N-5}}\
 \ldots\   x_5\  x_2\
 \begin{array}{||c||}
 \hline
 \hline
 {x_0}\\
 \hline
 \hline
 \ea
 \  x_3\  x_6\  \ldots\ {x_{N-4}}\  \begin{array}{|c|}
 \hline
 {x_{N-1}}\\
 \hline
 \ea  \\
 x_1\\
 x_4\\
 \vdots\\
 {x_{N-6}}\\
  \begin{array}{|c|}
 \hline
 {x_{N-3}}\\
 \hline
 \ea
 \ea
 \label{YRKL}
 \ee
This lattice connects the three (framed) endpoints with the central
(doubly framed) junction at $x_0$. The simplest version of a quantum
system living on this graph may/will employ again the RK
discretization of the kinetic energy (i.e., of the second derivative
operator, cf. Eq.~(\ref{SEdis})). The only exception is encountered
at $x_0$ where our choice of an acceptable matching is more flexible
(see Ref.~\cite{KS}). For the sake of simplicity we shall postulate
 \be
 -\frac{\psi(x_{1})+\psi(x_{2})+\cdots +\psi(x_{q})
 -u\,\psi(x_0)}{h^2}
 =E\,\psi(x_0)\,
 \label{SEdisq0}
 \ee
with a free parameter $u=u(q)$ set equal, say, to 3 at $q=3$. In the
bound-state arrangement this matching condition in the origin must
be complemented by the three ``asymptotic" Dirichlet boundary
conditions imposed at the remote ends of the edges.  The bound-state
energies will then coincide with the eigenvalues of the real and
symmetric $(3K+1)-$dimensional matrix Hamiltonian with partitioned
structure shown in Eq.~(\ref{blqpkk}). Wave functions will be
specified by Schr\"{o}dinger equation
 \be
  \left[ \begin {array}{ccccccc}
   3&-1&-1&-1&0&\ldots&0
  \\
  {}-1
&2&0&0&-1&\ddots&\vdots
 \\{}-1&0&2&0&0&\ddots&0
 \\{}-1&0&0&\ddots&\ddots&\ddots&-1
 \\
 {}0&-1&0
&\ddots&2&0&0
 \\{}\vdots&\ddots&\ddots&\ddots&0&2&0
 \\{}0&\ldots&0&-1&0&0&2\\
 \end {array} \right]\,
 \left[ \begin {array}{c}
 \psi(x_0)\\
 \psi(x_1)\\
 \psi(x_2)\\
 \psi(x_3)\\
 \vdots\\
 \psi(x_{N-2})\\
 \psi(x_{N-1})\\
 \ea
 \right ]=E\,
 \left[ \begin {array}{c}
 \psi(x_0)\\
 \psi(x_1)\\
 \psi(x_2)\\
 \psi(x_3)\\
 \vdots\\
 \psi(x_{N-2})\\
 \psi(x_{N-1})\\
 \ea
 \right ]\,.
 \label{kine}
 \ee
From the symmetry (i.e., Hermiticity) of the Hamiltonian one deduces
that at any integer $K$ the spectrum is real though not necessarily
nondegenerate. At $N=4$, for example, we get $E_{2,3}^{(4)}=2$ while
$E_{1,4}^{(4)}=5/2\mp \sqrt {13}/2$.

\subsection{X-shaped model and its star-shaped descendants with $q \geq 4$}

At $q=4$ the lattice-points should be numbered in the same manner as
above,
 \be
 \ba
  \begin{array}{|c|}
 \hline
 {x_{N-2}}\\
 \hline
 \ea
  \\
  {x_{N-6}}\\
 \vdots\\
 x_3\\
  \begin{array}{|c|}
 \hline
 {x_{N-3}}\\
 \hline
 \ea\ {x_{N-7}}\
 \ldots\   x_6\  x_2\
 \begin{array}{||c||}
 \hline
 \hline
 {x_0}\\
 \hline
 \hline
 \ea
 \  x_4\  x_8\  \ldots\ {x_{N-5}}\
  \begin{array}{|c|}
 \hline
 {x_{N-1}}\\
 \hline
 \ea
  \\
 x_1\\
 \vdots\\
 {x_{N-8}}\\
  \begin{array}{|c|}
 \hline
 {x_{N-4}}\\
 \hline
 \ea
 \ea
 \label{XRKL}
 \ee
The extension of this pattern to any positive integer $q$ is
obvious. In the corresponding Hamiltonian (\ref{blqpkk}) we may keep
the RK-discretization-related scalar parameter $u=u(q)$ variable or
equal to its ``maximum" $u(q)=q$ tractable as natural after
embedding of our graph into a sufficiently high-dimensional space.
We may add that at any $u$ the degeneracy of the spectrum will grow
with $q$. For illustration we may use the  model with the smallest
dimensions $N=N(q)=q+1$ where the energy eigenvalue $E=2$ proves
$(q-1)-$times degenerate. This is easily seen from Eq.~(\ref{kine})
and/or from its $q>3$ generalizations once we put there,
tentatively, $\psi(x_0)=0$. The whole set of equations then
degenerates to the single constraint $\sum_{j=1}^q\psi(x_j)=u-2$
with $q-1$ linearly independent eigenvector solutions.

At the two remaining unknown energies $E=E_{1,q+1}\neq 2$ we may
normalize $\psi(x_0)=1$ and eliminate $\psi(x_j)=1/(2-E)$ at all
$j>0$. We arrive at the elementary Bethe-ansatz-type quadratic
secular equation $q/(2-E)=u-E$ giving the two missing roots in
closed form,
 \be
 2\,E_{1,q+1}=2+u\mp \sqrt{(2-u)^2+4q}\,.
 \ee
This is the first nontrivial $q-$star-graph-spectrum formula which
is, of course, compatible with its above-mentioned special case
computed at $u(q)=3$ for $q=3$.

\section{Innovation: Two changes of perspective \label{secIV}}

The message delivered by the examples presented in preceding
sections can be summarized as a recommendation that the current
discrete square-well eigenvalue problem with $q=2$ can easily be
generalized to its $q-$pointed-star analogues with any integer
$q\geq 2$. Formally these models may be characterized by the
$N-$dimensional partitioned Hamiltonian matrices $H^{(N)}$ of
Eq.~(\ref{blqpkk}) where we set $N=qK+1$ and use $q-$dimensional row
vectors $\vec{v}=(-1,-1, \ldots,-1)$ and $\vec{0}=(0,0,\ldots,0)$
and $q-$dimensional unit matrices and null-matrices $I$ and $0$,
respectively. Of course, nothing really new emerges in such an
elementary constructive project which requires just a routine
application of the well known principles of quantum mechanics.

The situation becomes much more exciting when the purely kinetic
nature of the Hamiltonian of a quantum graph is enriched by an
interaction added, preferably, at the vertices. For us, this option
opened a way toward two generalizations which will be described in
what follows. In essence, they will be based on the thorough change

\begin{itemize}

 \item
 of the naively Hermitian nature of the interactions (we shall advocate
 here
 the transition from the usual real and symmetric interaction
 matrices $H^{(int)}$ to their asymmetric alternatives
 preserving the reality of the spectrum,
 cf.
 paragraph
 \ref{intver} below),

 \item
 of the naively realistic assumption of the strict locality of the
 models (this will represent a further development of the idea proposed in
 Ref.~\cite{fund} and briefly recalled
 in paragraph \ref{essen} below).

\end{itemize}

\subsection{${\cal PT}-$symmetric interactions at vertices \label{intver} }

One of the purposes of our present text is to enrich the picture of
dynamics of bound states living on quantum graphs via an
introduction of certain nontrivial interactions at their vertices.
In a broader physical context this is the project inspired not only
by Ref.~\cite{fund} (on bound states) but also by some of our other
papers (dealing with scattering). In the language of mathematics,
the formal connections between these two physical scenarios are
quite close, especially in the RK discretized models. Thus, although
there is no space here for a deeper study of the scattering on the
${\cal PT}-$symmetric graphs, we find it meaningful to mention,
briefly, at least some of the possible parallels.

\subsubsection{A brief detour to scattering models}

In our few recent papers on scattering
\cite{prd,discrete,smear,crypto} the introduction of certain
elementary nearest-neighbor ${\cal PT}-$symmetric interactions
between RK lattice points proved fruitful as a very useful and
productive model-building principle. Unfortunately,  there exist
several obstacles for making the analogy between the bound- and
scattering-state one-dimensional RK-based models sufficiently close.
Firstly,  one must keep in mind that in the scattering scenario the
number $N$ of the RK lattice points must be kept very large or
infinite. Secondly, the very essence of the arrangement of the
scattering experiments requires that the interactions themselves
should preferably be localized very close to the origin
\cite{prd,Jones}. In contrast, the bound-state arrangement of
Schr\"{o}dinger equations seems to prefer the transfer of the
support of interactions to the remote ends of the interval of
coordinates.  In such a case, perceivable technical simplifications
were reported not only in the one-dimensional continuous-coordinate
square-well models (cf. Refs.~\cite{David}) but also in the
realistic three-dimensional discrete-lattice calculations (cf.
Ref.~\cite{Molinari}).

This being said, a note on some lattice-based models of scattering
may still prove approprate. Firstly we could classify them more
easily in our present graph-based language. The presence of a
nearest-neighbor coupling will be indicated by the insertion of
symbol $\diamondsuit$ between the corresponding two lattice points.
In the  scattering-inspired arrangement these points are usually
chosen as lying not too far from the origin. In the first step the
following modification is obtained of the discrete graph of
Eq.~(\ref{VgRKL}),
 \be
 \ba
  \begin{array}{|c|}
 \hline
 {x_{2K-1}}\\
 \hline
 \ea \ {x_{2K-3}}\
 \ldots\   x_3\  x_1\ \diamondsuit\
 \begin{array}{||c||}
 \hline
 \hline
 {x_0}\\
 \hline
 \hline
 \ea
 \   x_2\  x_4\  \ldots\ {x_{2K-2}}\
   \begin{array}{|c|}
 \hline
 {x_{2K}}\\
 \hline
 \ea  \\
  \ea\,.
 \label{VgRKLdia}
 \ee
For illustration of a quantum system living on this graph let us
recall the interaction matrix of Ref.~\cite{prd},
 \be
  H^{(int)}=\left[ \begin {array}{cccc}
     0&g&0&\ldots
  \\
    -g&0&\ldots&
  \\{}0&\ldots&&
 \\
 \vdots&\ddots&&
 \end {array} \right]\,.
 \label{fora}
 \ee
This real and antisymmetric (i.e., ${\cal PT}-$symmetric \cite{prd})
matrix has to be added to the purely kinematic discrete square-well
Hamiltonian (\ref{blqpkk}). Unfortunately, some of the predictions
of this oversimplified model are unphysical \cite{Jonesdva}. In
subsequent Ref.~\cite{discrete} another version of ${\cal
PT}-$symmetric interaction has been proposed, therefore. It employed
the fully symmetrized localization of the nearest-neighbor
interactions in the RK graph,
 \be
 \ba
  \begin{array}{|c|}
 \hline
 {x_{2K-1}}\\
 \hline
 \ea \ {x_{2K-3}}\
 \ldots\   x_3\  x_1\ \diamondsuit\
 \begin{array}{||c||}
 \hline
 \hline
 {x_0}\\
 \hline
 \hline
 \ea
 \ \diamondsuit\  x_2\  x_4\  \ldots\ {x_{2K-2}}\
   \begin{array}{|c|}
 \hline
 {x_{2K}}\\
 \hline
 \ea  \\
  \ea\,
 \label{VgRKLdiabe}
 \ee
leading to the amended interaction matrix
 \be
  H^{(int)}=\left[ \begin {array}{ccccc}
     0&g&g&0&\ldots
  \\
    -g&0&\ldots&&
  \\-g&0&\ldots&&
  \\{}0&\ldots&\ddots&&
 \\
 \vdots&\ddots&&&
 \end {array} \right]\,.
 \ee
In Refs.~\cite{discrete} and \cite{crypto} we further shifted the
diamonds $\diamondsuit$ (representing the localization of
interactions) by one step in the lattice and arrived at the next
graph
 \be
 \ba
  \begin{array}{|c|}
 \hline
 {x_{2L-1}}\\
 \hline
 \ea\
 {x_{2L-3}}\
 \ldots\   x_3\ \diamondsuit\ x_1\
 \begin{array}{||c||}
 \hline
 \hline
 {x_0}\\
 \hline
 \hline
 \ea
 \  x_2\ \diamondsuit\   x_4\  \ldots\ {x_{2L-2}}\
 \begin{array}{|c|}
 \hline
 {x_{2L}}\\
 \hline
 \ea  \\
  \ea\,
 \label{V2RKL}
 \ee
yielding the next eligible interaction matrix
 \be
  H^{(int)}=\left[ \begin {array}{c|cc|cc|cc}
     0&0&0&0&\ldots&&
  \\
  \hline
    0&0&0&g&0&\ldots&
  \\0&0&0&0&g&0&\ldots
  \\
  \hline
  0&-g&0&0&0&0&\ldots
  \\0&0&-g&0&0&0&\ldots
  \\
  \hline
  0&0&0&0&0&0&\ddots
 \\
 \vdots&\vdots&\vdots&\vdots&\vdots&\ddots&\ddots
 \end {array} \right]\,.
 \label{desce}
 \ee
A general pattern emerges clearly. The whole class of interactions
can be realized via four nonvanishing matrix elements which are not
necessarily located just in the closest vicinity of the origin. This
quadruplet of off-diagonal matrix elements is allowed to move away
from the origin forming a series of descendants of
Eq.~(\ref{desce}).

The main benefit of this series of models defined on RK lattices is
threefold. Firstly, their study opens the way toward the {\em
unitary} scattering systems described by the sufficiently elementary
${\cal PT}-$symmetric Hamiltonians \cite{prd}. Secondly,  the
physical predictions (i.e., the reflection and transmission
coefficients) retain the form of closed formulas \cite{discrete}.
Thirdly, these models of scattering may find generalizations living
on some suitable classes of nontrivial quantum graphs in the nearest
future.

\subsubsection{${\cal PT}-$symmetric bound-state models with $q=2$}

In contrast to the scattering scenario where, typically, the matrix
in Eq.~(\ref{desce}) is infinite-dimensional, the RK version of the
bound-state problem may always be considered finite-dimensional.
Then, the repeatedly shifted symbol $\diamondsuit$ of the
interaction must ultimately reach the ends of the V-shaped graph,
 \be
 \ba
  \begin{array}{|c|}
 \hline
 {x_{2L-1}}\\
 \hline
 \ea\
 \diamondsuit\ {x_{2L-3}}\
 \ldots\   x_3\  x_1\
 \begin{array}{||c||}
 \hline
 \hline
 {x_0}\\
 \hline
 \hline
 \ea
 \  x_2\  x_4\  \ldots\ {x_{2L-2}}\
 \diamondsuit\  \begin{array}{|c|}
 \hline
 {x_{2L}}\\
 \hline
 \ea  \\
  \ea\,.
 \label{VRKL}
 \ee
The related exceptional Hamiltonian matrix represents the modified
square well with a nontrivial ${\cal PT}-$symmetric interaction
which is localized solely in the closest vicinity of the external
vertices. The related quantum Hamiltonian acquires the partitioned
$(K+1)-$dimensional tridiagonal form
  \be
 H=H^{(N)}(\lambda)=
 \left[
  \begin {array}{c|cccc|c}
 u&\vec{v}^T&0&\cdots&\cdots&0
 \\
 \hline
 \vec{v}&2I&-I&\ddots&\ddots&\vdots
 \\0&-I&\ddots&\ddots&\ddots&0
 \\\vdots&\ddots&\ddots&2I&-I&0
 \\
 \vdots&&\ddots&-I&2I&c(\lambda)
 \\
 \hline
 0&\cdots&\cdots&0&c(-\lambda)&2I
 \end {array} \right]\,
 \label{blockk}
 \ee
i.e., at $q=2$,
 \ben
 H= \left[ \begin {array}{c|cc|cc|cc|cc}
 \hline
 2&-1&-1&0&0&0\ \ &\multicolumn{2}{l}{\ \ \cdots}&0
 \\
  \multicolumn{6}{l}{\rule{5.3cm}{.1mm}\   \ldots}&
  \multicolumn{3}{r}{ \ldots\ \,
 \rule{1.7cm}{.1mm}}
   \\
    -1&2&0&-1&0&\multicolumn{2}{l}{\ \ \ \ 0  \ \ \ldots}&\cdots
   &0
 \\-1&0&2&0
&-1&\multicolumn{3}{l}{ \ \ \ddots \ \ \ \ddots
   \ \ \ \   }&\vdots
 \\
  \multicolumn{6}{l}{\rule{4.3cm}{.1mm} \ \ \ldots}&
  \multicolumn{1}{r}{ }&\ddots&\vdots
   \\
 0&-1&0&2&\multicolumn{2}{l}{\ \  0 \ \  \vdots\ \   \ddots}&
 \multicolumn{3}{r}{ \ldots\ \,
 \rule{4.1cm}{.1mm}}
 \\
 \vdots&
 \ddots&-1&\ddots&
 \multicolumn{2}{l}{\ddots \ \ \  \ \ \ddots}& \ddots&-1+{{}\lambda}&0
  \\
  \multicolumn{6}{l}{\rule{3.3cm}{.1mm} \ \ \ldots}&2
 &0&-1+{{}\lambda}
 \\
  \multicolumn{2}{l}{\ \ \  \ \ \vdots\ }&
  \multicolumn{7}{r}{\vdots\ \ \ \ \   \ \ldots \ \ \
 \rule{6.7cm}{.1mm}}
  \\
  \vdots&&\multicolumn{1}{c}{ }&&\ddots&
 -1-{{}\lambda}&0&2&0
 \\{}0&0&\multicolumn{1}{c}{\cdots }&&\cdots&0&-1-{{}\lambda}&0&2
 \\
 \hline
 \end {array} \right]\,.
 \een
The rightmost lowest corner carries all the dependence of the
Hamiltonian on the coupling (note that we changed its symbol from
$g$ to $\lambda$). This parallels the preferences recommended in
Refs.~\cite{David} or \cite{Molinari}.

\subsection{Introduction of nonlocality via inner products \label{essen}}

Whenever we declare a matrix [e.g., our Hamiltonian (\ref{blockk})
considered in the RK coordinate representation] manifestly
non-Hermitian, we almost always have in mind just the
non-Hermiticity in the current $\ell_2-$representation of the
Hilbert space. This space may be denoted  by the symbol ${\cal
H}^{(F)}$ where the superscript stands for the ``first" or
``friendly" space (cf. also \cite{SIGMA}). In this space the usual
formula
 \be
  \sum_{k=0}^{N-1} \psi^*_1(x_k) \psi_2(x_k)\ :=\br \psi_1|\psi_2\kt
 \label{jednac}
 \ee
defines the inner product between any pair of its elements (i.e.,
finite- or infinite-dimensional vectors) $\psi_1$ and $\psi_2$. In
this setting the authors of Ref.~\cite{Geyer} noticed and emphasized
that {\em the same} Hamiltonian may appear to be Hermitian in
another Hilbert spaces ${\cal H}^{(S)}$ where our choice of the
superscript stands for  the ``second" or ``subtle" space and where
{\em the same} set of vectors is merely assigned the following {\em
different}, non$-\ell_2$ inner product using a suitable nontrivial
``metric" $\Theta\neq I$,
 \be
  \sum_{j=0}^{N-1}  \sum_{k=0}^{N-1}  \psi^*_1(x_j)
 \Theta(x_j,x_k) \psi_2(x_k)=
 \br \psi_1|\Theta|\psi_2\kt \ :=
 \bbr \psi_1|\psi_2\kt \,.
 \label{laots}
 \ee
One can always make use of this flexibility of basic definitions,
keeping only in mind that the standard probabilistic
interpretation can {\em solely} be assigned to a Hamiltonian which
is Hermitian (in whatever Hilbert space). In this sense the models
described by asymmetric real Hamiltonian matrices with real
spectra {\em do not} leave the territory of the standard formalism
of quantum theory.

Although the latter idea has thoroughly been explained by several
authors \cite{Geyer,Carl,Alirev,DDTrev}, some of its key aspects
and consequences may be summarized in two brief sentences.
Firstly, we must assume that the spectrum of our Hamiltonians
$H=H(\lambda)$ remains real in some non-empty interval of the
measures of their asymmetry $\lambda$. In the second step we have
to introduce an invertible operator $\Omega$ which maps our
$\ell_2$ Hilbert space ${\cal H}^{(F)}$ onto another, unitarily
non-equivalent ``physical" $\ell_2$ Hilbert space ${\cal H}^{(P)}$
which is expected unitarily equivalent to the ``subtle" physical
space ${\cal H}^{(S)}$ endowed with nontrivial metric and product
(\ref{laots}).

More thoroughly, both these steps will be explained in section
\ref{aret}. Now let us only add that their practical appeal has
been well illustrated in nuclear physics where ${\cal H}^{(P)}$
represented the textbook Hilbert space of nucleons (i.e.,
fermions) while both the auxiliary Hilbert spaces ${\cal
H}^{(F,S)}$ were identified with the spaces of certain artificial,
effective ``interacting bosons" (cf. Ref.~\cite{Geyer} for more
details).

In Ref.~\cite{fund} we also worked with the triplet of spaces
${\cal H}^{(F,S,P)}$ and emphasized there the deep technical {\em
nontriviality} of the construction of the necessary metric
operator $\Theta=\Theta(H)$ in terms of which our asymmetric,
non-Hermitian real-matrix representations of the Hamiltonians were
made Hermitian with respect to the {\em ad hoc} inner product
(\ref{laots}). In particular, as long as we worked in coordinate
representation, we made distinction between the models which were
{\em local} (i.e., where $\Theta=\Theta_0(H) \neq I$ remained
represented by a diagonal matrix)  and {\em nonlocal}, i.e.,
characterized by the non-diagonal metrics
$\Theta_1(H)\,,\Theta_2(H)\,,\ldots$. Moreover, the most
unexpected property of the oversimplified models as studied in
Ref.~\cite{fund} has been revealed in the fact that, via a
suitable renumbering, one could achieve that the $j-$th metric
$\Theta_j(H)$ was represented by a very special $(2j+1)-$diagonal
matrix.

In our present paper we intend to demonstrate that these results may
be extended to a broad family of quantum graphs.

\subsubsection{A return to (hidden) Hermiticity of observables
\label{aret}}

In the formalism described in Ref.~\cite{Geyer} the  simple but
non-Hermitian Hamiltonian matrix $H\neq H^\dagger$ defined in ${\cal
H}^{(F)}$ has been put in correspondence with its idealized
isospectral partner $\mathfrak{h}=\mathfrak{h}^{(N)}(\lambda)$. The
latter operator is defined in ${\cal H}^{(P)}$ and it may be assumed
complicated. The correspondence is realized by the Dyson map,
 \be
 \Omega: H \to \mathfrak{h}=\Omega\,H\,\Omega^{-1}\,
 \label{oma}
 \ee
which is, by definition, non-unitary,
$\Omega=\Omega^{(N)}(\lambda)\neq \left (1/\Omega\right
)^\dagger$. Thus, we are allowed to require the Hermiticity  of
the isospectral partner Hamiltonian,
  \be
  \mathfrak{h}^{(N)}(\lambda)=\Omega(\lambda)\,H^{(N)}(\lambda)\
  \Omega^{-1}(\lambda)= \left [\mathfrak{h}^{(N)}\right ]^\dagger(\lambda)\,.
  \label{edi}
  \ee
The latter relation can be re-read as a constraint imposed upon the
simpler operator $H=H^{(N)}(\lambda)$ itself,
 \ben
 \Omega\,H^{(N)}\Omega^{-1}=\left [\Omega\,H^{(N)}\Omega^{-1}
 \right ]^\dagger =\left [\Omega^{-1} \right ]^\dagger
 \left [H^{(N)} \right ]^\dagger \Omega^\dagger\,.
 \een
In the re-arranged and abbreviated form this relation coincides with
the condition of a hidden  Hermiticity or ``quasi-Hermiticity"
\cite{Geyer,Dieudonne} of $H^{(N)}(\lambda)$,
 \be
 \left [H^{(N)} \right ]^\dagger
 =\Theta\,H^{(N)}\,\Theta^{-1}\,,
 \ \ \ \ \ \Theta=\Omega^\dagger\Omega >0
 \,.
 \label{cryptog}
  \ee
The closest correspondence between metric $\Theta$ of
Eq.~(\ref{laots}) and the Dyson map $\Omega$ is established in this
manner.

\subsubsection{The reconstruction of the {\em ad hoc} metric $\Theta=\Theta(H)$ \label{meto}}

In any ${\cal PT}-$symmetric quantum model, i.e., for Hamiltonians
with the property $H^\dagger={\cal P}\,H\,{\cal P}^{-1}$ ({\em and}
with the real spectrum) the correct physical probabilistic
interpretation of bound states must be based on the reconstruction
of the metric in ${\cal H}^{(S)}$. The matrix elements of this
metric may be made available as a solution of the linear algebraic
system of Eqs.~(\ref{cryptog}),
 \be
 \sum_{k=1}^N\,
 \left [
      \left (H^\dagger\right )_{jk}\,\Theta_{kn}
      -\Theta_{jk}\,H_{kn}\right ] =0
 \,,\ \ \ \ \ j,n=1,2,\ldots,N
   \,.
 \label{htottot}
 \ee
Needless to repeat that the resulting metrics are
Hamiltonian-dependent and by far not unique in general,
$\Theta=\Theta_j(H)$, $j=0, 1, \ldots$. Each of them defines a new,
independent Hermitian conjugation and, hence, a respective
independent $N-$dimensional physical Hilbert space ${\cal
H}^{(S)}\equiv {\cal H}^{(N)}_j$. The knowledge of the metric is
substantial. The parallel availability of the factor $\Omega$ and of
its conjugate $\Omega^\dagger$ remains less essential (though note
their role in section \ref{less} below). For this reason we shall
pay our main attention here just to the constructive assignment of
one or several alternative metrics $\Theta=\Theta^{(N)}_j(\lambda)$,
$j=0,1,\ldots$ to a given, ``prescribed" quantum-graph Hamiltonian
$H=H^{(N)}(\lambda)$.

This project consists of fulfilling two separate subtasks. Firstly,
we shall search for the metric in the form of a superposition
 \be
 \Theta=\Theta_{\beta_0,\beta_1,\ldots}^{(N)}=
  \beta_0\,{\cal P}_0^{(N)}+\beta_1\,{\cal P}_1^{(N)}+\ldots\,
  \label{enpar}
 \ee
of some suitable Hermitian, sufficiently simple though not
necessarily positive definite auxiliary components. Secondly, due
attention must be paid to the positive definiteness of the
metric~(\ref{enpar}) controlled by the appropriate choice of
parameters $\beta_j$. In addition, all of the ``pseudometrics"
${\cal P}_\mu^{(N)}$ will individually be assumed compatible with
the Hermiticity condition (\ref{cryptog}),
 \ben
 \sum_{k=1}^N\,
 \left [
      \left (H^\dagger\right )_{jk}\,\left ({\cal P}_\mu^{(N)} \right )_{kn}
      -\left ({\cal P}_\mu^{(N)} \right )_{jk}\,H_{kn}\right ] =0
 \,,\ \ \ \ \ j,n=1,2,\ldots,N,
 \een
 \be
 \ \ \ \ \ \ \ \ \ \ \ \ \ \ \ \ \
 \ \ \ \mu=0,1,\ldots
   \,.
 \label{htot}
 \ee
In this manner our ansatz (\ref{enpar}) will specify metrics
$\Theta$ as superpositions of pseudometrics ${\cal P}={\cal
P}^{(N)}={\cal P}^{(N)}_1, {\cal P}^{(N)}_2, \ldots$ which will be
required to possess a sparse-matrix structure. At $q=2$ this idea
has been shown productive in Ref.~\cite{fund}. In our present paper
we just extend this recipe to the ${\cal PT}-$symmetric quantum
graphs with $q \geq 3$.

\subsubsection{Nonlocal metrics: their sample construction at
$N=4$\label{dia4ane}}

For quantum systems living on the smallest Y-shaped discrete graph
 \be
 \ba
  \begin{array}{|c|}
 \hline
 {x_{2}}\\
 \hline
 \ea\,
 \begin{array}{||c||}
 \hline
 \hline
 {x_0}\\
 \hline
 \hline
 \ea\, \begin{array}{|c|}
 \hline
 {x_{3}}\\
 \hline
 \ea  \\
  \begin{array}{|c|}
 \hline
 {x_{1}}\\
 \hline
 \ea
 \ea
 \label{YRKL4}
 \ee
no space is left for the end-point additional interactions since we
do not wish that the matching point $x_0$ gets involved. Thus, our
present $N=4$ Y-shaped quantum graph will remain purely kinematic.
Its spectrum of energies will coincide with the eigenvalues of the
four-dimensional matrix Hamiltonian
 \be
 H^{(4)}(0)=\left[ \begin {array}{cccc} 3&-1&-1&-1
  \\-1&2&0&0
\\\noalign{\medskip}-1&0&2&0\\\noalign{\medskip}-1&0&0&2\end {array}
 \right]\,.
 \label{styr}
 \ee
The latter particular matrix is real, symmetric (i.e., Hermitian)
and positive definite. These properties (plus its natural
commutativity with itself) make this matrix eligible as an
admissible metric. Further metrics compatible with their implicit
algebraic definition~(\ref{cryptog}) can be sought as arbitrary
polynomial  functions of Hamiltonian (\ref{styr}),
 \be
 \Theta^{(4)} = \Theta^{(4)}_{\{c_0,c_1,\ldots\}}
 =
  c_0\,I+
  c_1\,H^{(4)}(0)+
  c_2\,\left [H^{(4)}(0)\right ]^2+\cdots\,.
 \label{twehle}
 \ee
The recipe is quick since the necessary explicit construction of the
integer powers of the Hamiltonian is straightforward yielding
 \be
 \Theta^{(4)}_{\{0,0,1,\ldots\}}
 =
\left [H^{(4)}(0)\right ]^2=
 \left[ \begin {array}{cccc} 12&-5&-5&-5\\\noalign{\medskip}-5&5&1&1
\\\noalign{\medskip}-5&1&5&1\\\noalign{\medskip}-5&1&1&5\end {array}
 \right]
 \label{stra}
 \ee
etc. Unfortunately, the construction of metrics via
Eq.~(\ref{twehle}) cannot be transferred to non-Hermitian matrices
$H \neq H^\dagger$. Another unpleasant feature of the metrics
sampled by Eq.~(\ref{stra}) lies in their non-sparse, full-matrix
form. For both of these reasons a return is recommended  to the
methods of paragraph \ref{meto}. Their results are universal --
for example, metric Eq.~(\ref{stra}) appears as a special case of
formula (\ref{ctyrka}) [cf. Section \ref{4.1} below] at $a=12$,
$b=-5$ and $f=j=k=1$.

\section{The proofs of the reality of energies \label{reali}}


At any integer number $q$ the energy spectrum of our quantum graphs
is partially degenerate at $\lambda=0$. This leaves the
specification of a complete basis ambiguous. Another ambiguity
emerges via the non-Dirac metrics $\Theta_j \neq I$, $j=0,1,\ldots$.
We may construct several alternative, {\em nonequivalent}
representations of the respective ``correct" or ``selected" Hilbert
space of states ${\cal H}^{(N)}({\lambda})={\cal
H}^{(N)}_j({\lambda})$, $j=0,1,\ldots$.  In the respective inner
products (\ref{laots}) one encounters mutually nonequivalent metrics
$\Theta^{(N)}_j({\lambda})$ sampled in paragraph~\ref{dia4ane}
above. The discovery of such a new freedom of making the choice
between alternative inner products can be perceived as belonging to
the most important recent achievements in quantum physics, with
impact ranging from the new flexibility of the interacting boson
models in nuclear physics \cite{Geyer} and from formulations of
several new theoretical ideas in quantum mechanics \cite{Alirev} up
to the emergence of the new classes of phenomenological Lagrangians
in quantum field theory \cite{Carl} where, e.g., the presence of
ghosts can successfully be eliminated in some cases \cite{Chen} and
where even the concept of integrability acquired an updated meaning
\cite{DDTrev}. The use of the varying non-Dirac metrics $\Theta \neq
I$ also opened the way toward new challenges connected, e.g., with
the description of bound states in time-dependent systems
\cite{timedep} or in the relativistic kinematical regime
\cite{alikg}. In some phenomenological models of scattering the
variability of $\Theta$ has  been suggested as a guarantee of the
causality and/or unitarity of the process
\cite{prd,discrete,Jones,Jonesdva}.

In our present treatment of the ambiguity of $\Theta=\Theta_j$,
$j=0,1,\ldots$ we shall be guided by the approach of paper
\cite{fund}. We considered there the standard coordinate
representation $\br x| \Theta |x'\kt$ of the metric operator and
required that a suitable measure of its ``nonlocality" [i.e., of its
deviation from the Dirac's ``local" delta function $\delta(x-x')$]
should be identified with the postulate of the existence of
fundamental length $\theta$ which characterizes the physical system
in question. The same philosophy will also be accepted in our
present text. We shall assume that the appeal of the concept of
fundamental length survives the transition to the discrete-graph
Hamiltonians of any dimension $N =qK+1$ with $K\geq 1$. We feel that
partially nonlocal models with nonvanishing elementary lengths might
find a very natural area of applicability in quantum graphs since
the experimental waveguides and other nanotechnological realizations
of quantum graphs almost certainly contain an uncertainty in the
localization related to the degree of idealization of the real
physical system in question \cite{Milos}. In addition, the
observability of the coordinate in a quantum graph may prove
overridden by the transfer of emphasis to some other measured
quantities (cf., e.g., the famous question ``Can one hear the shape
of a graph?" as asked in Ref.~\cite{[24]}).

\subsection{Numerical proofs }

We believe that even the oversimplified discrete quantum graphs
with not too large $q$ and/or $N$ can offer a new source of
entirely abstract elementary models with, say, an unusual or
anomalous parameter-dependence of their spectra tractable by
numerical techniques. Having this purely descriptive ambition in
mind let us now study the first few $q=3$ models in some detail,
emphasizing that a key to all of the above-sampled applications of
non-Dirac metrics $\Theta \neq I$ lies in the demonstration of the
reality of the spectrum of the initial Hamiltonian $H$ which is
non-Hermitian, $H \neq H^\dagger$ in ${\cal H}^{(F)}$.

\subsubsection{The Y-shaped discrete quantum graph with $N=7$
\label{lokarno}}

%
\begin{figure}[h]                     
\begin{center}                         
\epsfig{file=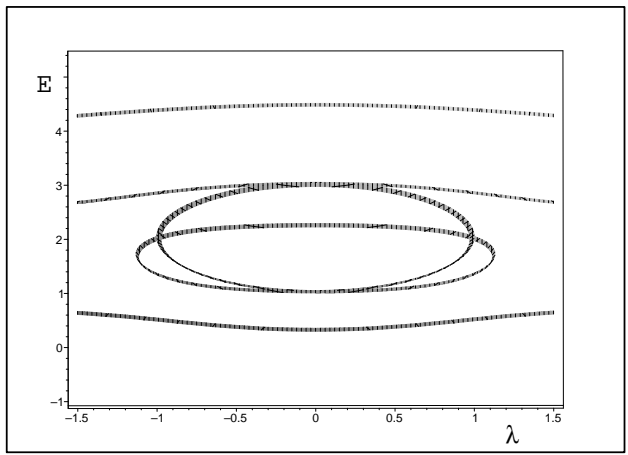,angle=270,width=0.6\textwidth}
\end{center}                         
\vspace{-2mm} \caption{The spectrum of $H^{(7)}(\lambda)$.
 \label{fionej}}
\end{figure}

 \noindent
The first nontrivial discrete $q=3$ graph
 \be
 \ba
  \begin{array}{|c|}
 \hline
 {x_{5}}\\
 \hline
 \ea \diamondsuit  {x_{2}}\
 \begin{array}{||c||}
 \hline
 \hline
 {x_0}\\
 \hline
 \hline
 \ea
 \  {x_{3}}  \diamondsuit    \begin{array}{|c|}
 \hline
 {x_{6}}\\
 \hline
 \ea  \\
 {x_{1}}\\
  \begin{array}{|c|}
 \hline
 {x_{4}}\\
 \hline
 \ea
 \ea
 \label{YRKL7}
 \ee
leads to the seven-dimensional one-parametric family of Hamiltonians
 \be
 H^{(7)}= \left[ \begin {array}{c|ccc|ccc}
 3&-1&-1&-1&0&0&0
   \\
  \hline
  {}-1
 &2&0&0&-1&0&0
 \\{}-1&0&2&0&0&-1+{{}{\lambda}}&0
 \\{}-1&0&0&2&0&0&-1-{{}{\lambda}}
 \\
  \hline{}0&-1&0
 &0&2&0&0
 \\{}0&0&-1-{{}{\lambda}}&0&0&2&0
 \\{}0&0&0&-1+{{}{\lambda}}&0&0&2
  \end {array} \right]\,.
 \label{model7}
 \ee
They exhibit a particularly tight mutual interaction between the
endpoints. The analysis of the energy spectrum pertaining to
$H^{(7)}(\lambda)$ may rely on the construction of the secular
polynomial which appears factorized into its quadratic and quintic
component. Thus, two of the levels are prescribed by explicit
formulae, $E_{2,5}=2\mp\sqrt {1-{{\it {\lambda}}}^{2}}$, while the
remaining ones follow from the reduced secular equation,
 \ben
{{{}{E}}}^{5}-11\,{{{}{E}}}^{4}+ \left( {{{}{\lambda}}}^{2}+43
\right) {{{}{E}}}^{3}- \left( 7\,{{{}{\lambda}}}^{2}+72 \right)
{{{}{E}}}^{2}+
 \left( 14\,{{{}{\lambda}}}^{2}+48 \right) {{}{E}}-7\,{{\it
 {\lambda}}}^{2}-9=0\,.
 \een
This confirms that the energy levels of our seven-point Y-shaped
quantum graph remain real in the interval of couplings ${\lambda}
\in (-1,1)$. Its endpoints coincide with the position of the Kato's
``exceptional points", i.e., of the values at which the first merger
and complexification of a pair of energies takes place.

The overall $\lambda-$dependence of energies is displayed in Figure
\ref{fionej}. We see there that the spectrum has four fragile (i.e.,
asymptotically complex) and three robust (i.e., never complexifying)
components. This observation fits the pattern predicted by the
generic tunable ${\cal PT}-$symmetric model of Ref.~\cite{fragile}.

\subsubsection{The next, $q=3$ model with $N=10$}

%
\begin{figure}[h]                     
\begin{center}                         
\epsfig{file=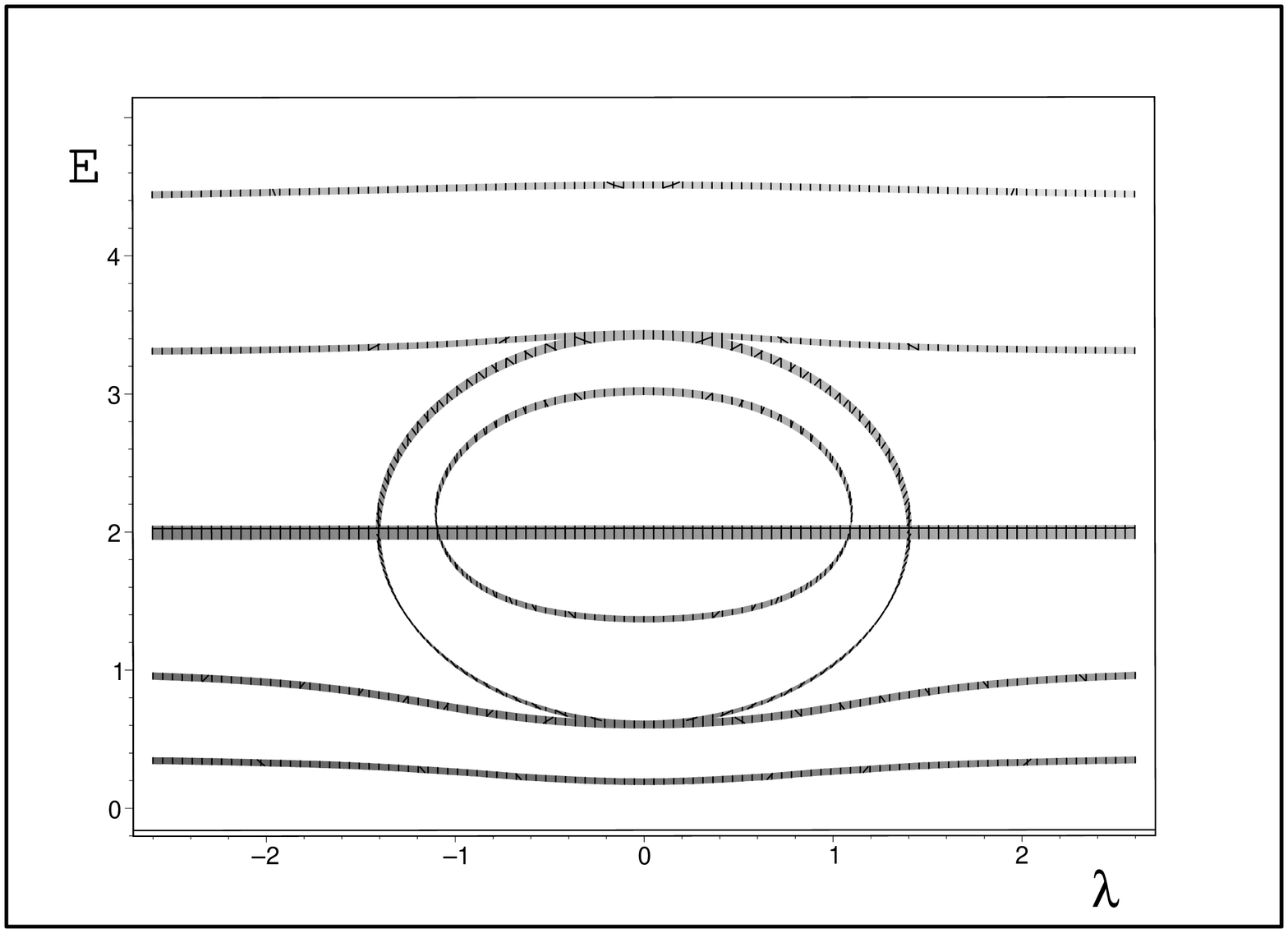,angle=270,width=0.6\textwidth}
\end{center}                         
\vspace{-2mm} \caption{The spectrum of $H^{(10)}(\lambda)$.
 \label{fibj}}
\end{figure}

On the ten-point graph-lattice
 \be
 \ba
  \begin{array}{|c|}
 \hline
 {x_{8}}\\
 \hline
 \ea \diamondsuit {x_{5}}\
  x_2\
 \begin{array}{||c||}
 \hline
 \hline
 {x_0}\\
 \hline
 \hline
 \ea
 \  x_3\     {x_{6}} \diamondsuit  \begin{array}{|c|}
 \hline
 {x_{9}}\\
 \hline
 \ea  \\
 x_1\\
  {x_{4}}\\
  \begin{array}{|c|}
 \hline
 {x_{7}}\\
 \hline
 \ea
 \ea
 \label{YRKL10}
 \ee
our Hamiltonian $H^{(10)}(\lambda)$ acquires the matrix form
 \ben
  \left[ \begin {array}{c|ccc|ccc|ccc}
 3&-1&-1&-1&0&0&0&0&0&0
 \\
  \hline
   -1&2&0&0&-1&0&0&0&0&0
 \\{}-1&0&2&0&0
&-1&0&0&0&0
 \\{}-1&0&0&2&0&0&-1&0&0&0
 \\
 \hline
 0&-1&0&0&2&0&0&-1&0&0
 \\{}0&0&-1&0&0
&2&0&0&-1+{{}{\lambda}}&0
 \\{}0&0&0&-1&0&0&2&0&0&-1-{{}{\lambda}}
 \\
 \hline
 0&0&0&0&-1&0&0&2&0&0
 \\{}
 0&0&0&0&0&
 -1-{{}{\lambda}}&0&0&2&0
 \\{}0&0&0&0&0&0&-1+{{}{\lambda}}&0&0&2
 \end {array} \right]\,.
 \een
The set of its eigenvalues comprises the constant and doubly
degenerate doublet $E_{5,6}=2$, the two explicit roots
$E_{3,8}=2\pm\sqrt {2-{{{}{\lambda}}}^{2}}$ and the six implicit
nodal zeros of the reduced secular polynomial
 \ben
 {{{}{E}}}^{6}-13\,{{{}{E}}}^{5}+ \left( {{{}{\lambda}}}^{2}+63
 \right) {{{}{E}}}^{4}- \left( 9\,{{{}{\lambda}}}^{2}+140 \right)
 {{{}{E}}}^{3} +\left( 25\,{{{}{\lambda}}}^{2}+141
 \right) {{\it {E}}}^{2}-
 \een
  \ben \ \ \ \
  -\left(22\,{ {{}{\lambda}}}^{2}+56 \right)
 {\it {E}}+5\,{{{}{\lambda}}}^{2}+6 =0.
 \een
The $\lambda-$dependence of these energies is displayed in
Figure~\ref{fibj} where the thickness of the middle straight line
emphasizes that the exceptional constant-energy level $E=2$ is
doubly degenerate.

\subsection{Nonnumerical proof}

The most straightforward rigorous proof of the reality of the
energies [i.e., of the reality of the spectrum of Hamiltonian
$H^{(N)}(\lambda)$] may proceed via the explicit constructive
demonstration of existence of at least one metric
$\Theta=\Theta(H)\neq I$ which makes this Hamiltonian Hermitian in
${\cal H}^{(S)}$.

\subsubsection{The local versions of the discrete quantum graphs}

For our  Y-shaped graphs the dimension $N=3K+1$ is finite so that we
may search for special solution $\Theta_0$ of Eq.~(\ref{htottot})
using a diagonal matrix ansatz and some computer-assisted symbolic
manipulations. In this way we verified that at $N=7$ the diagonal
solution is positive definite and, up to an overall factor, unique,
 \ben
 \Theta_{(diagonal)}^{(7)}=
  \left[ \begin {array}{c|ccc|ccc}
         1&0&0&0&0&0&0
   \\
   \hline
       0&1&0 &0&0&0&0
 \\{}0&0&1&0&0&0&0
 \\{}0&0&0&1&0  &0&0
 \\
 \hline
 0&0&0&0&1&0&0
 \\{}0&0&0&0&0&{ \frac {1-{\it  {\lambda}}}{1+{{}{\lambda}}}}&0
 \\{}0&0&0&0&0&0&{ \frac {1+{\it  {\lambda}}}{1-{{}{\lambda}}}}
 \end {array} \right]\,.
 \een
At any $N=3K+1$ with $K=3,4,\ldots$ we then revealed that the
verification of the absence of any non-diagonal elements in the
difference $\Theta_{(diagonal)}^{(7)}-I$ can be performed
non-numerically. Finally, using the assumption of diagonality we
reduced the matrix difference $H^\dagger\Theta-\Theta\,H$ in
Eq.~(\ref{htottot}) to the mere pair of equations which specified
the last two missing matrix elements in our ultimate solution
compatible with Eq.~(\ref{htottot}) at any integer $K$,
 \be
 \Theta_{(diagonal)}^{(3K+1)}=
  \left[ \begin {array}{cccccc}
   1&0&0&\cdots&\cdots&0
  \\0&1&0 &{\ddots}&{}&\vdots
 \\\vdots&{\ddots}&\ddots&{\ddots}&{}&
 \\{}{}&{}&{}&1&{\ddots}&\vdots
 \\ \vdots&{}&{}&{\ddots}& { \frac {1-{\it  {\lambda}}}{1+{{}{\lambda}}}}&0
 \\{0}&{\cdots}&{}&{\cdots}&{0}& { \frac {1+{\it {\lambda}}}{1-{{}{\lambda}}}}
 \end {array} \right]\,.
 \label{diagora}
 \ee
Obviously, this matrix is invertible, Hermitian and positive
definite so that it may play the role of the metric inside the whole
interval of couplings ${\lambda}\in (-1,1)$. This confirms that our
Hamiltonian $H^{(N)}(\lambda)$ becomes Hermitian in the {\em ad hoc}
Hilbert space ${\cal H}^{(S)}$ where the diagonal metric
(\ref{diagora}) is employed. Thus, we may modify our notation, write
$\Theta_{(diagonal)}^{(3K+1)}=\Theta_0^{(N)}(\lambda)$ and ${\cal
H}^{(S)} \ \equiv\ {\cal H}_0^{(N)}(\lambda)$ and re-read the latter
statement as the rigorous proof of the reality of the energies for
${\lambda}\in (-1,1)$.

\subsubsection{Equivalent Hermitian Hamiltonians \label{less}}

Our constructive proof of existence of the (unique) diagonal metric
$\Theta=\Theta_0^{(N)}(\lambda)$ given by Eq.~(\ref{diagora})
implies the survival of the observability of the RK coordinates in
both our (unitarily equivalent) physical Hilbert spaces ${\cal
H}^{(P)}$ and, in an amended notation, ${\cal H}^{(S)} \ \equiv\
{\cal H}_0^{(N)}(\lambda)$. As long as the diagonality of
$\Theta_0^{(N)}(\lambda)$ is specified in coordinate representation,
the usual multiplicative operator of coordinates remains Hermitian
in the {\em same} Hilbert space ${\cal H}_0^{(N)}(\lambda)$ as the
Hamiltonian $H^{(N)}(\lambda)$, indeed.

In such an exceptional case it makes sense to recollect the
Dyson-mapping-related factorization $\Theta=\Omega^\dagger\Omega$ of
our diagonal metric and to restrict our attention to the positive
definite and diagonal operator factors $\Omega = \sqrt{\Theta}$.
Their knowledge enables us to recall definition (\ref{edi}) and, for
illustration, to evaluate the related Hermitian isospectral partner
Hamiltonian $\mathfrak{h}=\mathfrak{h}^{(N)}(\lambda)$, say, at
$N=7$,
 \ben
 \mathfrak{h}=\left[ \begin {array}{ccccccc}
   3&-1&-1&-1&0&0&0
   \\{}-1
&2&0&0&-1&0&0
   \\{}-1&0&2&0&0&-\sqrt { 1-{\it
{\lambda}}^2}  &0
    \\{}-1&0&0 &2&0&0&-\sqrt { 1-{\it
{\lambda}}^2}
    \\{}0&-1&0&0&2&0&0
    \\{}0&0&
-\sqrt { 1-{{}{\lambda}}^2} &0 &0&2&0
   \\{}0&0&0&-\sqrt { 1-{\it
{\lambda}}^2} &0&0&2
\end {array} \right].
 \een
The block-tridiagonal generalization of this formula to all
dimensions $N=3K+1$ is obvious.

\section{Manifestly nonlocal quantum graphs \label{blokybe}}

Our general quantization recipe described in paragraph \ref{essen}
admits the transition from the diagonal metric $\Theta_0$
[exemplified by Eq.~(\ref{diagora}) at $q=3$] to its arbitrary
non-equivalent alternative (\ref{enpar}). This means that in the
spirit of trivial examples studied in Ref.~\cite{fund} we are
allowed to violate the locality also in all the other ${\cal
PT}-$symmetric quantum graphs. Moreover, we can demand that the
sequence of the not necessarily unique nondiagonal metrics
$\Theta_1$, $\Theta_2$ $\ldots$ is partially ordered with respect to
their increasing degree of non-locality defined, in a way suggested
in Ref.~\cite{fund}, as a suitable growing function
$\theta=\theta_j$ of subscript $j$ which is proportional, say, to
the number of nonzero diagonals in the matrices or metrics
$\Theta_j$.

Expansion (\ref{enpar}) of each individual  $\Theta_j$ combines, in
principle, several indefinite pseudometrics ${\cal P}={\cal P}(H)$.
In this sense, our main task is twofold: Firstly we have to find at
least one solution of Eq.~(\ref{htot}), the nonlocality of which
saturates the number $\theta_j$. Secondly we must guarantee the
positivity of the resulting multiparametric sum $\Theta_j$.

\subsection{Sparse-matrix pseudometrics  at $N=4$\label{4.1} }

For our model (\ref{styr}) the brute-force solution of
Eq.~(\ref{cryptog}) leads  to the following most general and
exhaustive five-parametric formula for the (pseudo)metric,
 \be
 \Theta^{(4)}(a,b,f,j,k)= \left[ \begin {array}{cccc}
a&b&b&b\\\noalign{\medskip}b&b-f-j+a&f&j
\\\noalign{\medskip}b&f&b-f-k+a&k\\\noalign{\medskip}b&j&k&b-j-k+a
\end {array} \right]\,.
 \label{ctyrka}
 \ee
The variability of one of the parameters is spurious since it merely
signals the double degeneracy of one of the eigenvalues. The fifth
degree of freedom may immediately be interpreted, therefore, as an
inessential angle of rotation in the two-dimensional subspace
spanned by the corresponding pair of eigenvectors. The remaining
four real parameters are independent and their presence reflects the
well known ambiguity of the assignment of the metric $\Theta$ to a
given Hamiltonian (for a thorough discussion of this mathematical
subtlety with serious physical consequences cf., e.g.,
Ref.~\cite{Geyer}).

\subsubsection{Positive-definite cases (metrics)}

The apparent simplicity of formula (\ref{ctyrka}) is slightly
misleading because the interpretation of the matrix
$\Theta^{(4)}(a,b,f,j,k)$ as a metric requires that we guarantee its
positive definite status~\cite{Geyer}. At $N=4$, this property would
be equivalent to the positivity of all of its four eigenvalues
$\tau_j>0$, $j=0,1,2,3$,
 \ben
 \tau_{0,1}= a+b/2\pm b\,\sqrt {13}\,/2\,,
 \ \
  \ \
 \een
 \be
 \ \ \ \
 \tau_{2,3}= a+b-f-j-k \pm\sqrt
 {{f}^{2}+{j}^{2}+{k}^{2}-fj-fk-kj}\,.
 \label{posi}
 \ee
We see that the positivity of the metric (i.e., of the norm) is
guaranteed by the specification of the allowed domain ${\cal D}$  of
our quintuplet of parameters. This is particularly easy when we
restrict our attention to the subdomain of ${\cal D}$ where
$f=j=k=0$ and where
 \ben
  \Theta^{(4)}(a,b,0,0,0)=\left[ \begin {array}{cccc} a&b&b&b
  \\\noalign{\medskip}b&b+a&0&0
\\\noalign{\medskip}b&0&b+a&0\\\noalign{\medskip}b&0&0&b+a\end {array}
 \right]\,.
 \een
We obtain the complete positivity constraint
 \ben
  2\,a>|b|\,(\sqrt {13}+1)\,,
 \ \
  \ \ b \leq  0\,,
 \een
 \ben
  2\,a>b\,(\sqrt {13}-1)\,,
 \ \
  \ \ b >  0\,.
 \een
This means that the allowed values of $b$ belong to an interval
which grows with $a>0$. As long as this guarantees the positivity of
$\tau_{0,1}$ at all $f,j,k$, the specification of the allowed domain
of parameters will be completed by the inequality
 \be
  f+j+k +\sqrt
 {{f}^{2}+{j}^{2}+{k}^{2}-fj-fk-kj}<a+b\,.
 \ee
Out of the doublet of constraints $\tau_{2,3}>0$ this is equivalent
to the stronger one. We see that neither of the three parameters
$f,j,k$ will be allowed to get too large in comparison with $a$.

\subsubsection{Indefinite cases (generalized parities ${\cal P}$) \label{loukot}}

Elementary $N=4$ example looks particularly well suited for
illustrative purposes. Thus, in a search for the simplest possible
{\em parity-type} pseudometrics ${\cal P}^{(4)}(a,b,f,j,k)$ we have
to construct such a solution of Eq.~(\ref{htot}) which {\em is}
invertible but which {\em is not} positive definite. This matrix may
be represented by the same formula as the metric
$\Theta^{(4)}(a,b,f,j,k)$ but at least one of the positivity
constraints (\ref{posi}) must be violated. In such a setting the
requirement of maximal simplicity may start from the elimination of
$b$ which is exceptional in occurring nine times in
Eq.~(\ref{ctyrka}). At $b=0$ we may also normalize $a=1$ [${\cal
P}^{(4)}(a,0,f,j,k)$ wouldn't be invertible at $a=0$] and have
 \be
 {\cal P}^{(4)}(1,0,f,j,k)= \left[ \begin {array}{cccc}
1&0&0&0\\\noalign{\medskip}0&1-f-j&f&j
\\\noalign{\medskip}0&f&1-f-k&k\\\noalign{\medskip}0&j&k&1-j-k
\end {array} \right]\,.
 \label{ctyrka1}
 \ee
For a maximal simplicity of this matrix we leave just one of its
parameters nonzero and get, say,
 \be
 {\cal P}^{(4)}(1,0,0,0,1)= \left[ \begin {array}{cccc}
1&0&0&0\\\noalign{\medskip}0&1&0&0
\\\noalign{\medskip}0&0&0&1\\\noalign{\medskip}0&0&1&0
\end {array} \right]\,.
 \label{ctyrka12}
 \ee
In ${\cal PT}-$symmetric models this  matrix can play the role of
parity ${\cal P}$. Geometrically, it realizes the left-right
reflection of our Y-shaped graph (\ref{YRKL4}).

\subsection{Block-tridiagonal pseudometrics }

\subsubsection{Solutions of Eq.~(\ref{htot}) at $N=7$ and $N=10$}

The main source of insight in the structure of metrics and
pseudometrics lies in the natural partitioning of Hamiltonians
$H^{(N)}$ in  $q-$dimensional submatrices. At $q=3$  the first two
nontrivial though still sufficiently sparse matrix solutions were
obtained by the straightforward computer-assisted symbolic
manipulations with Eq.~(\ref{htot}). The results
 \be
 {\cal P}^{(7)}_1(\lambda)=\left[ \begin {array}{c|ccc|ccc}
 -1&1&1&1&0&0&0\\
 \hline
 1&0&0&0&1&0&0
 \\{}1&0&0&0&0&1-{{}{\lambda}}&0
 \\{}
 1&0&0&0&0&0&1+{{}{\lambda}}
 \\
 \hline
 0&1&0&0&0&0&0
 \\{}0&0&1-{{}{\lambda}}&0&0&0&0
 \\{}0&0&0&1+
 {{}{\lambda}}&0&0&0\\
 \end {array} \right]\,
 \label{firno}
 \ee
and
 \be
 {\cal P}^{(10)}_1(\lambda)= \left[ \begin {array}{c|ccc|ccc|ccc}
   -1&1&1&1&0&0&0&0&0&0
 \\
 \hline
 {}1&0&0&0&1&0&0&0&0&0
  \\
  {}1&0&0&0&0&1
&0&0&0&0
 \\
 {}1&0&0&0&0&0&1&0&0&0
  \\
  \hline
  {}0&1
&0&0&0&0&0&1&0&0
 \\
 {}0&0&1&0&0&0&0&0&1-{{}{\lambda}}&0
 \\
 {}0&0&0&1&0&0&0&0&0&1+{{}{\lambda}}
  \\
  \hline
  {}0&0
&0&0&1&0&0&0&0&0
 \\
 {}0&0&0&0&0&1-{{}{\lambda}}&0&0&0&0
 \\
 {}0&0&0&0&0&0&1+{
\it {\lambda}}&0&0&0\\
\end {array} \right]\,
 \label{firnobe}
 \ee
open the way toward  extrapolations.

\subsubsection{Extrapolation to any $N=3K+1$}

The knowledge of the nontrivial solutions (\ref{firno}) and
(\ref{firnobe}) of Eq.~(\ref{htot}) inspires the proposal of the
following block-partitioned ansatz
  \be
 {\cal P}^{(3K+1)}_1(\lambda)=
 \left[
  \begin {array}{c|cccc|c}
 w&\vec{v}^T&0&\cdots&\cdots&0
 \\
 \hline
 \vec{v}&0&-I&\ddots&\ddots&\vdots
 \\0&-I&\ddots&\ddots&\ddots&0
 \\\vdots&\ddots&\ddots&0&-I&0
 \\
 \vdots&&\ddots&-I&0&d(\lambda)
 \\
 \hline
 0&\cdots&\cdots&0&d(\lambda)&0
 \end {array} \right]\,.
 \label{blockkse}
 \ee
Using the Hamilotnian of Eq.~(\ref{blockk}) in its $q=3$ version
 \ben
  \left[ \begin {array}{c|ccc|ccc|ccc}
 3&-1&-1&-1&0&0&\multicolumn{2}{l}{\ \ \cdots}&\cdots&0
 \\
  \multicolumn{7}{l}{\rule{6.1cm}{.1mm}\ \  \ldots}&
  \multicolumn{3}{r}{ \ldots\ \,
 \rule{1.7cm}{.1mm}}
  \\
    -1&2&0&0&-1&0&\multicolumn{3}{l}{\ \ \ddots}
   &\vdots
 \\-1&0&2&0&0
&-1&\multicolumn{2}{c}{ \ \ \ddots
   \ \ \ \   \vdots \ \ \  \vdots}&&
 \\-1&0&0&2&0&\ddots&\ddots&0&\vdots&\vdots
 \\
  \multicolumn{6}{l}{\rule{4cm}{.1mm} \ \ \ldots}&
  \multicolumn{4}{l}{ \ldots\ \,
 \rule{4.7cm}{.1mm}}
  \\
 0&-1&0&0&2&\ddots&\ddots&-1&0&0
 \\
 \vdots&
 0&-1&0&
 0 &\ddots&\ddots&0&-1+{{}{\lambda}}&0
 \\
 \multicolumn{2}{c}{ \vdots }&\ddots&\multicolumn{2}{c}{\ddots \vdots
 \ddots}
 &\ddots&\ddots&0&0&-1-{{}{\lambda}}\vspace{-.21cm}
 \\
  \multicolumn{2}{l}{}&
 \multicolumn{8}{l}{\ldots \ \,
 \rule{9cm}{.1mm}}
  \\
  \multicolumn{2}{c}{ \vdots }&\cdots&0&-1&0&0&2&0&0
 \\
  \vdots&&&\cdots&0&
 -1-{{}{\lambda}}&0&0&2&0
 \\{}0&\cdots&&\cdots&0&0&-1+{{}{\lambda}}&0&0&2
 \end {array} \right]
 \een
(were only the right low corner is coupling-dependent) and
performing the appropriate insertions one readily verifies that
Eq.~(\ref{htot}) becomes an identity provided  only that the unknown
submatrix $d(\lambda)$ is defined by the elementary formula
$d(\lambda)=-c(\lambda)$ which remains the same at all integers
$K=2,3,\ldots$.

Naturally, our block-tridiagonal ansatz (\ref{blockkse}) as well
as its verification and subsequent conclusions may immediately be
extended to the other star graphs with $q=4,5,\ldots$. The details
are left to the reader. In what follows we shall address, instead,
the other two questions, viz., a transition from the
block-tridiagonal pseudometrics (\ref{blockkse}) to their
block-pentadiagonal and higher descendants (cf. paragraph
\ref{pentagram} below) and a transition from the indefinite
pseudometric matrices [exemplified here by Eq.~(\ref{blockkse})]
to the acceptable and positive definite band-matrix metrics
expressed by the first nonlocal formula
 \be
 \Theta=\Theta_{[\beta]}^{(N)}=
  \beta\,{\Theta}_0^{(N)}+{\cal P}_1^{(N)}\,
  \label{supe2}
 \ee
[cf. paragraph \ref{sepere} below and note that the latter
expression is just the two-term truncated version of the general
expansion~(\ref{enpar})].

\subsection{Block-pentadiagonal pseudometrics \label{pentagram}}

The appeal of finding a block-pentadiagonal pseudometric (denoted by
the symbol ${\cal P}_2^{(N)}$ here) would lie in its possible
insertion in the next truncated version of formula (\ref{enpar}),
 \be
 \Theta=\Theta_{[\beta,\gamma]}^{(N)}=
  \beta\,{\Theta}_0^{(N)}+\gamma\,{\cal P}_1^{(N)}+{\cal P}_2^{(N)}\,.
  \label{enparbe}
 \ee
This formula may be used to define the more smeared,
block-pentadiagonal nonlocal metrics.

Once we leave the positivity questions aside and choose $N=10$, the
application of the computer-assisted direct-solution algorithm
produces the  pseudometric solution ${\cal P}^{(10)}_2(\lambda)$ of
Eq.~(\ref{htot}) in the form
 \ben
 \left[ \begin {array}{c|ccc|ccc|ccc}
  0&0&0&0&1&1&1&0&0&0
 \\
 \hline
 {}0&-1&1&1&1&0&0&1&0&0
 \\
 {}0&1&-1&1&0
&1&0&0&1-{{}{\lambda}}&0\\{}0&1&1&-1&0&0&1&0&0&1+{{}{\lambda}}
 \\
 \hline
 {}1&1&0&0&-1&0&0&1&0&0
 \\
 {}1&0&1&0&0&
-{{{}{\lambda}}}^{2}-1&0&0&1-{{}{\lambda}}&0
 \\
 {}1&0&0&1&0&0&-{{ \it {\lambda}}}^{2}-1&0&0&1+{\it
{\lambda}}
 \\
 \hline
 {}0&1&0&0&1&0&0&-2&0&0
 \\
 {}0&0&1-{{}{\lambda}}&0&0&1-{{}{\lambda}}&0&0&{\frac {-2+2\,{
\it {\lambda}}-{{{}{\lambda}}}^{2}+{{{}{\lambda}}}^{3}}{1+{\it
{\lambda}}}}&0
 \\
 {}0&0&0&1+{{}{\lambda}}&0&0&1+{{}{\lambda}}&0&0&{\frac {-2-2\,
{{}{\lambda}}-{{{}{\lambda}}}^{2}-{{{}{\lambda}}}^{3}}{1-{\it
{\lambda}}}}
\end {array}
 \right]\,.
 \een
In a way recommended at $q=2$ in Ref.~\cite{fund}) this $q=3$
solution has been made unique by the requirement  of having a
minimum of nonvanishing matrix elements in the first row. In our
present case the optimality of such a requirement is less obvious.
Indeed, as long as we have to optimize the sum (\ref{enparbe})
rather than its individual components we may feel dissatisfied by
the comparatively high number (= 39) of nonvanishing matrix
elements in ${\cal P}^{(10)}_2(\lambda)$ [and, among them, 12
manifestly $\lambda-$dependent items]. In such a case we may
contemplate ${\cal P}^{(10)}_{2a}(\lambda)$ given by the formula
 \ben
 \left[ \begin {array}{c|ccc|ccc|ccc}
  2&-1&-1&-1&1&1&1&0&0&0
 \\
 \hline
 {}-1&0&1&1&0&0&0&1&0&0
 \\
 {}-1&1&0&1&0
&0&0&0&1-{{}{\lambda}}&0
 \\
 {}-1&1&1&0&0&0&0&0&0&1+{{}{\lambda}}
 \\
 \hline
 {}1&0&0&0&0&0&0&0&0&0
 \\
 {}1&0&0&0&0&
-{{{}{\lambda}}}^{2}&0&0&0&0
 \\
 {}1&0&0&0&0&0&-{{ \it {\lambda}}}^{2}&0&0&0
 \\
 \hline
 {}0&1&0&0&0&0&0&-1&0&0
 \\
 {}0&0&1-{{}{\lambda}}&0&0&0&0&0&{\frac {-1+{
\it {\lambda}}-{{{}{\lambda}}}^{2}+{{{}{\lambda}}}^{3}}{1+{\it
{\lambda}}}}&0
 \\
 {}0&0&0&1+{{}{\lambda}}&0&0&0&0&0&{\frac {-1-
{{}{\lambda}}-{{{}{\lambda}}}^{2}-{{{}{\lambda}}}^{3}}{1-{\it
{\lambda}}}}
\end {array}
 \right].
 \een
containing just a minimum -- 30 pieces -- of the nonvanishing matrix
elements. We may also ask for the absence of fractions at a cost of
having 32 nonvanishing matrix elements in ${\cal
P}^{(10)}_{2b}(\lambda)=$
 \ben
 = \left[ \begin {array}{c|ccc|ccc|ccc}
 3+{\lambda}^2&-1&-1&-1&1&1&1&0&0&0
 \\
 \hline
 {}-1&1+{\lambda}^2&1&1&0&0&0&1&0&0
 \\
 {}-1&1&1+{\lambda}^2&1&0
&0&0&0&1-{{}{\lambda}}&0
 \\
 {}-1&1&1&1+{\lambda}^2&0&0&0&0&0&1+{{}{\lambda}}
 \\
 \hline
 {}1&0&0&0&1+{\lambda}^2&0&0&0&0&0
 \\
 {}1&0&0&0&0&
 1&0&0&0&0
 \\
 {}1&0&0&0&0&0&1&0&0&0
 \\
 \hline
 {}0&1&0&0&0&0&0&{\lambda}^2&0&0
 \\
 {}0&0&1-{{}{\lambda}}&0&0&0&0&0&0&0
 \\
 {}0&0&0&1+{{}{\lambda}}&0&0&0&0&0&0
\end {array}
 \right].
 \een
Another option could be based of the compromising choice of ${\cal
P}^{(10)}_{2c}(\lambda)=$
 \ben
 =\left[ \begin {array}{c|ccc|ccc|ccc}
 3&-1&-1&-1&1&1&1&0&0&0
 \\
 \hline
 {}-1&1&1&1&0&0&0&1&0&0
 \\
 {}-1&1&1&1&0
&0&0&0&1-{{}{\lambda}}&0
 \\
 {}-1&1&1&1&0&0&0&0&0&1+{{}{\lambda}}
 \\
 \hline
 {}1&0&0&0&1&0&0&0&0&0
 \\
 {}1&0&0&0&0&
 1-{\lambda}^2&0&0&0&0
 \\
 {}1&0&0&0&0&0&1-{\lambda}^2&0&0&0
 \\
 \hline
 {}0&1&0&0&0&0&0&0&0&0
 \\
 {}0&0&1-{{}{\lambda}}&0&0&0&0&0&-{\lambda}^2\,\frac{1-{\lambda}}{1+{\lambda}}&0
 \\
 {}0&0&0&1+{{}{\lambda}}&0&0&0&0&0&-{\lambda}^2\,\frac{1+{\lambda}}{1-{\lambda}}
\end {array}
 \right]
 \een
with some elementary fractions but with minimum (= 8 pieces) of
$\lambda-$dependent matrix elements.

\subsection{Metrics as positive definite superpositions of pseudometrics \label{sepere}}

\subsubsection{Positivity constraint at $ N=7$}

The results of preceding paragraph have to be complemented by the
empirical observation that the candidate for the metrics which is
chosen in the one-parametric block-tridiagonal form
$\Theta_{[\beta]}^{(N)}=\beta \times \Theta_{0}^{(N)}+{\cal
P}_1^{(N)}$ need not necessarily be positive definite. This may
numerically be confirmed not only at vanishing $\beta=0$ but also at
the positive values of $\beta$ which are not sufficiently large. For
illustration we selected $\beta=1/10$ and found that in dependence
on the value of $\lambda$, three or four eigenvalues of
$\Theta_{[1/10]}^{(7)}(\lambda)$ remained negative.

%
\begin{figure}[h]                     
\begin{center}                         
\epsfig{file=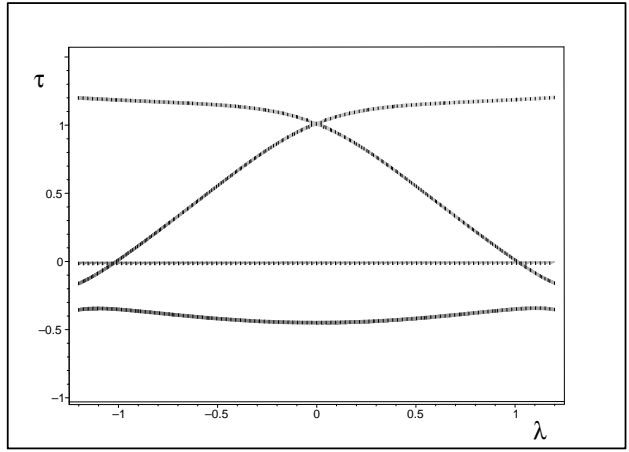,angle=270,width=0.6\textwidth}
\end{center}                         
\vspace{-2mm} \caption{The three lowest eigenvalues of the matrix $2
\, {\Theta}_0^{(7)}+{\cal P}_1^{(7)}$.
  \label{fiszxej}}
\end{figure}

One must be careful even if the candidate matrix
$\Theta_{[\beta]}^{(7)}(\lambda)$ looks dominated by its diagonal
and safely positive-definite metric component. This is illustrated
in Figure~\ref{fiszxej} where we displayed the three lowest
eigenvalues of $\Theta_{\beta}^{(N)}$ at $\beta=2$. In the picture
we also see that there already exists just single eigenvalue which
breaks the positivity and stays negative inside the whole interval
of $\lambda\in (-1,1)$.

%
\begin{figure}[h]                     
\begin{center}                         
\epsfig{file=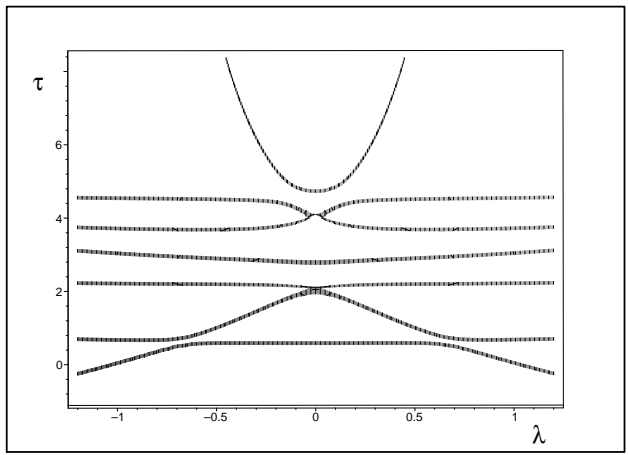,angle=270,width=0.6\textwidth}
\end{center}                         
\vspace{-2mm} \caption{The spectrum of the metric
$\Theta_{[3]}^{(7)}=3 \, {\Theta}_0^{(7)}+{\cal P}_1^{(7)}$.
 \label{fibesej}}
\end{figure}

Our subsequent Figure~\ref{fibesej} illustrates the situation in
which $\beta=3$ is sufficiently large. Similar pictures can offer
a comparatively reliable graphical confirmation of the positivity
of any candidate (\ref{enpar}) for the metric. Thus, in our
particular illustration we see that for
$\Theta_{[\beta]}^{(7)}(\lambda)$ considered in the whole interval
of ${\lambda} \in \left
(-{\lambda}^{(numerical)}(\beta),{\lambda}^{(numerical)}(\beta)\right
)$ it is sufficient to choose $\beta=3$. Then our picture also
leads to the graphical estimate of
${\lambda}^{(numerical)}(3)\approx 1$. As long as the dimension
$N=7$ is small, this estimate may be replaced by the rigorous
identification of ${\lambda}^{(numerical)}(3)= 1$. By means of
elementary algebra it is easy to show that this value coincides
not only with the singularity (i.e., with the point of divergence)
of the maximal eigenvalue of $\Theta_{[3]}^{(7)}(\lambda)$ but
also with the zero of the minimal eigenvalue of the same matrix.

%
\begin{figure}[h]                     
\begin{center}                         
\epsfig{file=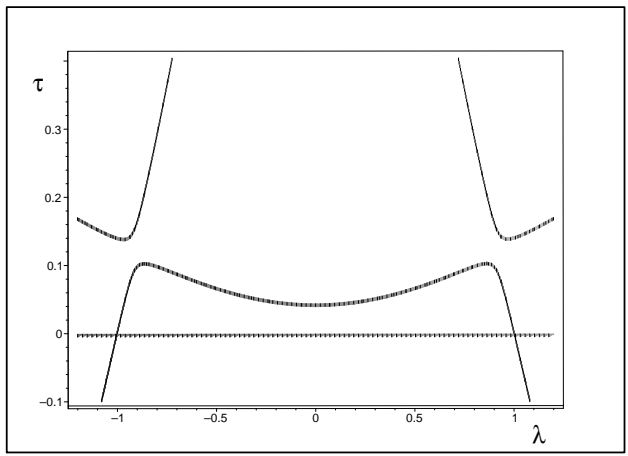,angle=270,width=0.6\textwidth}
\end{center}                         
\vspace{-2mm} \caption{The lowest two numerical eigenvalues of
metric $\Theta_{[5/2]}^{(7)}(\lambda)$
 \label{fzxej}}
\end{figure}

The $\lambda-$dependence of the minimal eigenvalue of
$\Theta_{[\beta]}^{(7)}(\lambda)$ is of particular relevance in the
regime where the value of $\beta$ decreases below it value used in
Figure~\ref{fibesej}. In our next Figure~\ref{fzxej} we use
$\beta=5/2$ and see that the loss of the positive-definiteness of
$\Theta_{[\beta]}^{(7)}(\lambda)$ may be expected to occur at
$\lambda=0$ where the lowest eigenvalue would vanish. Of course, as
long as the dimension of our illustrative model is small, it is very
easy to find the corresonding critical value of
 \be
 \beta_{minimal}=\frac{1}{3}+\frac{1}{6}\,\sqrt [3]{44+36\,i\sqrt {107}}
 +{\frac {26}{3}}\,{\frac {1}{
\sqrt [3]{44+36\,i\sqrt {107}}}}\,. \label{criti}
 \ee
This quantity lies, in rational arithmetics, inside interval
$(39/16,5/2)$ and is numerically approximated by $\sim 2.46050487$.
In a few complementary tests we found that the positivity of the
metrics  $\Theta_{[\beta]}^{(7)}(\lambda)$ is still reliably
confirmed at $\beta=149/60 \sim 2.483333333$ since in the standard
precision of computer  arithmetics the related minimum $\sim 0.02$
of the lowest numerical eigenvalue is still safely positive at
$\lambda=0$.

%
\begin{figure}[h]                     
\begin{center}                         
\epsfig{file=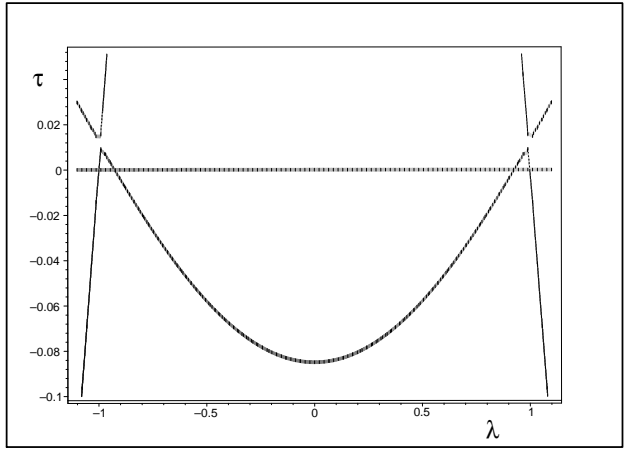,angle=270,width=0.6\textwidth}
\end{center}                         
\vspace{-2mm} \caption{The lowest two numerical eigenvalues of
matrix $\Theta_{[38/16]}^{(7)}(\lambda)$
 \label{fzej}}
\end{figure}

Matrix $\Theta_{[\beta]}^{(7)}(\lambda)$ retains its applicability
as a metric also for $\beta$s which lie slightly below their
universal, $\lambda-$independent bound (\ref{criti}). In these cases
one must restrict the admissible variability of the coupling
$\lambda$ to  intervals  ${\lambda} \in \left (-1,-{\lambda}^{(ad\
hoc)}(\beta)\right )$ and  ${\lambda} \in \left ({\lambda}^{(ad\
hoc)}(\beta),1\right )$. For explicit  numerical illustration of
such a conditional, $\lambda-$dependent positivity of the metric
below the critical boundary (\ref{criti}) we choose
$\beta=39/16=2.4375$ and revealed that the lowest numerical
eigenvalue of $\Theta$ (with the minimum $\sim -0.02<0$ at
$\lambda=0$) remained negative in the interval of ${\lambda} \in
(-{\lambda}^{(ad\ hoc)}, {\lambda}^{(ad\ hoc)})$ where
$\lambda^{(ad\ hoc)} \sim 0.5$ at our sample value of $\beta=
39/16$. It is necessary to keep in mind that ${\lambda}^{(ad\
hoc)}(\beta)$ quickly converges to one with the decrease of
$\beta<\beta_{minimal}$. This is well illustrated by our last
Figure~\ref{fzej} where we obtained $\lambda^{(ad\ hoc)} \sim 0.94$
at $\beta=38/16=2.375$.

\subsubsection{The positivity constraint at $ N=3K+1\geq 10$}

For the ten-dimensional metric candidates for the metric with the
same block-tridiagonal structure, e.g., for $\beta=3$ in
 \ben
 \Theta_{[3]}^{(10)}(\lambda)=
  \left[ \begin {array}{c|ccc|ccc|ccc}
  2&1&1&1&0&0&0&0&0&0
\\
\hline
 1&3&0&0&1&0&0&0&0&0
 \\{}1&0&3&0&0&1
 &0&0&0&0
 \\{}1&0&0&3&0&0&1&0&0&0
 \\
 \hline
 0&1
 &0&0&3&0&0&1&0&0
 \\{}0&0&1&0&0&3&0&0&1-{{}{\lambda}}&0
 \\{}0&0&0&1&0&0&3&0&0&1+{{}{\lambda}}\\
 \hline
 0&0
&0&0&1&0&0&3&0&0\\{}0&0&0&0&0&1-{{}{\lambda}}&0&0&3\,{ \frac {1-{\it
{\lambda}}}{1+{{}{\lambda}}}}&0\\{}0&0&0&0&0&0&1+{ \it
{\lambda}}&0&0&3\,{\frac {1+{{}{\lambda}}}{1-{{}{\lambda}}}}
\end {array} \right]\,
 \een
very similar results were obtained  supporting the applicability of
our above-presented considerations to all the sequence of metric
candidates $\Theta_{[\beta]}^{(N)}(\lambda)$ with sufficiently large
parameters  $\beta>\beta_{minimal}(N)$ and with unconstrained
dimensions $N=3K+1 = 13, 16,\ldots$.

\section{Discussion and summary\label{finis}}

In a way emphasized by several authors \cite{Jones,cubic} one
should, strictly speaking, distinguish between the $x-$dependence in
the wave function $\psi(x)$ and the $x-$dependence in the potential
$V(x)$ since in these two functions the concept of locality has a
different mathematical background as well as physical meaning.
Usually, the variable $x$ entering wave functions $\psi(x)$ is
treated as a measurable (i.e., real) quantity while the choice of
the local $V(x)$ may be treated just as a very special case of its
possible generalized, equally admissible non-local alternatives.

The more widespread use of the ``non-local" wave functions $\psi(x)$
(where $x$ need not be an observable real coordinate) only occurred
during the growth of popularity of differential-operator
Hamiltonians $H_{(PT)}=p^2+V_{(PT)}(x)$ where $x$ has been
considered complex \cite{Carl}. A not too dissimilar non-locality
also characterizes our present nonlocal versions of ${\cal
PT}-$symmetric discrete quantum graphs where we left the physical
meaning of the spatial coordinate unspecified, citing only the
related thorough discussion of this question available in our
preceding paper Ref.~\cite{fund}.

We may summarize our present results by saying that we transferred
the concept of ${\cal PT}-$symmetry to the class of quantum
systems living on graphs. These graphs generalize the usual real
line of coordinates in one dimension. Several non-Hermitian,
${\cal PT}-$symmetric versions of these structures have been
studied. On technical level we found one of the most vital
mathematical sources of encouragement in a few older papers
\cite{Katka} whose authors demonstrated the practical viability of
an approximative reduction of the graph edges to discrete lattices
of points and {\em vice versa}. On this background we succeeded in
combining the existing quantum-graph concepts with the very fresh
formalism using ${\cal PT}-$symmetric Hamiltonians which are only
made Hermitian via a comparatively complicated {\em ad hoc} inner
product.

The technical feasibility of such a synthesis had several
independent reasons. First of all, the spectra of energies proved
real for the range of couplings $\lambda$ which stayed independent
of the changes of the dimension $N$ of the lattice. Second, our
choice of the model proved lucky in the sense that in the current
coordinate basis one of the {\em constructed} metric matrices
$\Theta\neq I$ happened to remain strictly diagonal. Together with
the elementary form of matrix elements of this particular metric
$\Theta=\Theta_0$ this not quite expected result made the necessary
rigorous proof of the reality of the energies virtually trivial.

As the first byproduct of this circumstance one of the eligible
physical interpretations of our apparently non-Hermitian
quantum-graph system remained trivial in the sense that it just
required an inessential modification of the concept of observables
and that it enabled us to construct its spectrally equivalent
representation characterized by the Hamiltonian which is Hermitian
in the current sense (cf. operator $\mathfrak{h}$ in paragraph
\ref{less} above).

In the same theoretical framework the second important consequence
of the existence of the well-defined interval of admissible
couplings may be seen in the emergence of new freedom in the
choice of alternative, {\em different} physical interpretations of
the same quantum-graph Hamiltonian $H(\lambda)$. We were, once
more, lucky in revealing that there exists an extremely natural
partial ordering of these interpretations dictated merely by the
degree of their nonlocality or, in other words, by the extent of
the smearing of the coordinate (the degree of this smearing or, if
you wish, fundamental length $\theta_j$) grew with the subscript
$j$ of the selected closed-form metric $\Theta_j$).

On descriptive side let us re-emphasize the minimality of our
interactions which were not supported by the whole graph but just by
the closest vicinity of its endpoints. This also contributed to the
feasibility of our constructions for which we had to develop several
computer-assisted auxiliary symbolic-manipulation techniques and
adaptive algorithms. Fortunately, the explicit calculations which
were performed at the smallest dimensions usually generated the
output which admitted an extrapolation. Hence, the subsequent
adaptation of the algorithms often degenerated to the mere
verification of the extrapolated ansatz.

During these constructions we completely avoided the unnecessarily
complicated direct construction of the non-diagonal Dyson-map
matrices $\Omega$ and restricted our attention just to the metrics.
Moreover we revealed that these metrics can be decomposed into sums
of certain sparse pseudometrics, i.e., matrices with a sufficiently
large portion of matrix elements equal to zero. This facilitated our
calculations  at higher dimensions.

Our requirement of a fixed nonlocality does not make the resulting
metric $\Theta=\Theta(H)$ unique. The constructive analysis of this
metric-ambiguity problem in the specific quantum-graph setting can
be perceived as an extension of several recent non-graph (or
trivial-graph, $q=2$) studies assigning several non-equivalent
probabilistic interpretations to a given Hamiltonian \cite{Dibwe}. A
partial correspondence can be then seen to the standard transitions
between the coordinate and momentum representations of wave
functions $\psi(x)$ where the role of the (unitary) Fourier
transformation of Hilbert space is being taken over by the
manifestly non-unitary Dyson mapping $\Omega$. In such a setting we
made use of the fact that the argument $x$ of wave functions need
not necessarily carry the direct physical meaning of an (arbitrarily
precisely observable) coordinate. In terms of measurements the
immediate connection between the coordinate $x$ and its
observability is, therefore, weakened. The coordinates can be
interpreted as ``smeared" \cite{smear}. In the context of
topologically nontrivial quantum graphs the prospective utilization
of such a feature of phenomenological models looks particularly
promising.

In the language of mathematics our present family of discrete
quantum-graph models proved exceptionally friendly. Their choice
enabled us to disentangle the hidden Hermiticity constraints
(\ref{cryptog}) and to find closed formulae for the sparse-matrix
metrics. The resulting availability of their generic multiparametric
forms has been interpreted as a new freedom of a
phenomenology-friendly choice among alternative inner products
specifying the non-equivalent physical Hilbert spaces of states
${\cal H}^{(N)}_j$, $j=0,1,\ldots$. Whenever asked for, an extension
of our present particular quantitative and illustrative results on
${\cal PT}-$symmetric quantum graphs to the higher degrees of
nonlocality and/or beyond their equilateral $q-$point-star special
class with small $q=3,4,\ldots$ looks comparatively easy and
straightforward.

In the context of physics our present results are unexpectedly
encouraging. A new flexibility of the model-building has been
achieved here, first of all, via extension of the class of eligible
interactions and, secondly,  via the related innovative control of a
degree of nonlocality reflected by the introduction of the
``tunable" inner products. A deeper investigation of these
possibilities seems to form a new and promising
quasi-Hermitian-graph project filling a certain gap in the broader
context of existing directions of the study of quantum theory on
graphs.

\subsection*{Acknowledgements}

The support by the Institutional Research Plan AV0Z10480505, by the
M\v{S}MT ``Doppler Institute" project LC06002 and by GA\v{C}R grant
Nr. 202/07/1307 is acknowledged.



\begin{thebibliography}{00}


\bibitem{fund}
M. Znojil,
Phys. Rev. D 80, 045022 (2009).



\bibitem{Seba}
P. Exner and P. \v{S}eba, 
J. Math. Phys. 30, 2574 (1989).

\bibitem{Krejcirik}
%
T. Ekholm, H. Kova\v{r}\'{\i}k, and D. Krej\v{c}i\v{r}\'{\i}k, 
Arch. Ration. Mech. Anal. 188, 245 (2008).


\bibitem{napohtal}
illustrative pictures are available on address

http://en.wikipedia.org/wiki/Quantum\_graph

\bibitem{fractals}
V. Nekrashevych and A. Teplyaev, ``Groups and analysis on
fractals", review paper in proceedings \cite{exproc}.
%

\bibitem{exproc}
P. Exner,  J. P. Keating,  P. Kuchment,  and A. Teplyaev, Analysis
on Graphs and Its Applications (AMS, Rhode Island, 2008).
%
%
%
%
%
%
%
%
%


\bibitem{Kuchment}
P. Kuchment, 
Waves in Random Media 14, S107 (2004)
and
%
J. Phys. A: Math. Gen. 38, 4887 (2005);

%
Quantum graphs: an introduction and a brief survey, review paper in
proceedings \cite{exproc}, p. 291.
%




\bibitem{Exner}
P. Exner,  Leaky Quantum Graphs: A Review, review paper in
proceedings \cite{exproc}, p. 523.
%


\bibitem{Ragou}
E. Ragoucy, J. Phys. A: Math. Theor. 42, 295205 (2009).
%
%
%
%
%
%
%
%


\bibitem{Hatano3}
N. M. Shnerb, and D. R. Nelson, Phys. Rev. Lett. 80, 5172 (1998);

J. Feinberg, and A. Zee,  Phys. Rev. E 59, 6433 (1999),

R. A. Janik, M. A. Nowak, G. Papp, and I. Zahed, Acta Phys. Polon. B
30, 45 (1999);

J. Heinrichs, Phys. Rev. B 63, 165108  (2001).

\bibitem{Brody}
D. C. Brody, A. C. T. Gustavsson, and L. P. Hughston, Phys. A: Math.
Theor. 42, 295303 (2009).
%
%
%
%
%
%
%
%

\bibitem{Smil}
T. Kottos, and U. Smilansky, Phys. Rev. Lett. 79, 4794 (1997);

S. Gnutzmann,  and U. Smilansky, 
Advances in Physics 55, 527 (2006);

U. Smilansky, 
J. Phys. A.: Math. Theor. 40, F621 (2007).


\bibitem{Huill}
T. Huillet,  J. Phys. A: Math. Theor. 42, 275001 (2009).
%
%
%
%
%
%
%
%
%
%
%
%


\bibitem{Molinari}
L. G. Molinari, J. Phys. A: Math. Theor. 42, 265204 ( 2009).
%


\bibitem{Hatano}
N. Hatano, and D. R. Nelson, Phys. Rev. Lett. 77, 570 (1996).

\bibitem{Hatano2}
J. Feinberg, and A. Zee, Nucl. Phys. B 504, 579 (1997).

\bibitem{Geyer}
F. G. Scholtz, H. B. Geyer  and F. J. W. Hahne, Ann. Phys. (NY) 213,
74 (1992).

\bibitem{Kleefeld}
C. M. Bender and K. A. Milton, Phys. Rev. D 55, R3255  (1997) and
57, 3595 (1998);

%
C. M. Bender, D. C. Brody and H. F. Jones, Phys. Rev. Lett. 89,
270401 (2002);


M. C. Ogilvie, and P. N. Meisinger, SIGMA 5, 047 (2009).


\bibitem{Andrianov}
A. A. Andrianov, F. Cannata, and A. Y. Kamenshchik, J. Phys. A:
Math. Gen. 39, 9975 (2006).

\bibitem{BB}
C. M. Bender and S. Boettcher, Phys. Rev. Lett. 80, 5243 (1998).


\bibitem{proc}
Proceedings of the Workshop, ``Pseudo-Hermitian Hamiltonians in
Quantum Physics," Prague, June 2003, Ed. by M. Znojil, Czech. J.
Phys. 54 (2004), No. 1, dedicated issue;

Proceedings of the Workshop, ``Pseudo-Hermitian Hamiltonians in
Quantum Physics II," Prague, June 2004, Ed. by M. Znojil, Czech. J.
Phys. 54 (2004), No. 10, dedicated issue;

Proceedings of the Workshop, ``Pseudo-Hermitian Hamiltonians in
Quantum Physics III," Istanbul, June 2005, Ed. by M. Znojil, Czech.
J. Phys. 55 (2005), No. 9, dedicated issue;

Proceedings of the Workshop, ``Pseudo-Hermitian Hamiltonians in
Quantum Physics IV," Stellenbosch, November 2005, Ed. by H. Geyer,
D. Heiss and M. Znojil, J. Phys. A: Math. Gen. 39 (2006), No. 32,
dedicated issue;

Proceedings of the Workshop, ``Pseudo-Hermitian Hamiltonians in
Quantum Physics V," Bologna, July 2006, Ed. by M. Znojil, Czech. J.
Phys. 56 (2006), No. 9, dedicated issue;

Proceedings of the Workshop, ``Pseudo-Hermitian Hamiltonians in
Quantum Physics VI," London, July 2007, Ed. by A. Fring, H. Jones
and M. Znojil, J. Phys. A: Math. Theor. 41 (2008), No. 24, dedicated
issue;

Proceedings of the Workshop, ``Pseudo-Hermitian Hamiltonians in
Quantum Physics VII," Benasque, July 2008, Ed. by A. Andrianov et
al, SIGMA,  Vol. 5 (2009),  dedicated issue;

Proceedings of the VIIIth Conference, ``Non-Hermitian Hamiltonians
in Quantum Physics," Mumbai, January 2009, Ed. by S. R. Jain and Z.
Ahmed, Pramana - J.Phys. 73 (2009), No 2, dedicated issue.


\bibitem{Uwe}
U. Guenther, and
%
B. F. Samsonov,
Phys. Rev. A 78, 042115  (2008) and

%
Phys. Rev. Lett. 101, 230404 (2008).

\bibitem{alispectralsing}
A. Mostafazadeh, Pramana - J. Phys. 73, 269 (2009).


\bibitem{optics}
Z. H. Musslimani et al, 
Phys. Rev. Lett. 100, 030402 (2008);

M. V. Berry,  J. Phys. A: Math. Theor. 41,  244007 (2008);

O. Bendix, R. Fleischmann,  T. Kottos, and B. Shapiro, Phys. Rev.
Lett. 103, 030402 (2009).
%
%

\bibitem{Bijan}
B. Bagchi, and A. Fring, arXiv:0907.5354v1 [hep-th].


\bibitem{Tretter}
H. Langer, and C. Tretter, Czechosl. J. Phys. 54, 1113 ((2004).

\bibitem{BBjmp}
C. M. Bender, S. Boettcher,  and P. N. Meisinger, J. Math. Phys. 40,
2201 (1999).

\bibitem{erratum}
C. M. Bender,  D. C. Brody and H. F. Jones,
Phys. Rev. Lett. 
 92, 119902  (2004).


\bibitem{ptsqw}
M. Znojil, 
Phys. Lett. A 285, 7  (2001);

M. Znojil, and G. L\'{e}vai, 
Mod. Phys. Lett. A 16, 2273  (2001);

B. Bagchi, S. Mallik, and C. Quesne,
%
Mod. Phys. Lett. A 17, 1651 (2002);

B. Bagchi, H. B\'{\i}la, V. Jakubsk\'{y}, S. Mallik, C. Quesne,
and M.
Znojil, 
%
Int. J. Mod. Phys. A  
21, 2173 (2006).

\bibitem{Batal}
A. Mostafazadeh and A. Batal,  J. Phys. A: Math. Gen. 37, 11645
 (2004).
%

\bibitem{Weigert}
S. Weigert,  Czech. J. Phys. 55, 1183 (2005);

%
M. Znojil, 
J. Phys. A: Math. Gen. 39, 10247  (2006);

%
%
%
E. Ergun, SIGMA 5, 007 (2009).


\bibitem{KS}
V. Kostrykin, R. Schrader, 
J. Phys. A: Math. Gen. 32, 595 (1999).

\bibitem{prd}
M. Znojil,
Phys. Rev. D 78,  025026 (2008).


\bibitem{discrete}
M. Znojil,
J. Phys. A: Math. Theor. 41,  292002 (2008).


\bibitem{smear}
M. Znojil,
Phys. Rev. D 80, 045009 (2009).


\bibitem{crypto}
M. Znojil,  SIGMA 5, 085 (2009).
%


\bibitem{Jones}
H. F. Jones, Phys. Rev. D 76,  125003 (2007).
%
%

\bibitem{David}
D. Krej\v{c}i\v{r}\'{\i}k, H. B\'{\i}la  and M. Znojil,
J. Phys. A: Math. Gen. 39, 10143 (2006);

%
D. Krej\v{c}i\v{r}\'{\i}k,
J. Phys. A: Math. Theor. 41,  244012 (2008).

\bibitem{Jonesdva}
%
%
H. F. Jones,
Phys. Rev. D 78,  065032 (2008).


\bibitem{SIGMA}
M. Znojil,
SIGMA 5,  001 (2009).

\bibitem{Carl}
C. M. Bender,  
Rep. Prog. Phys. 70, 947 (2007).

\bibitem{Alirev}
A. Mostafazadeh, J. Math. Phys. 43, 205 and 2814 (2002);

A. Mostafazadeh, Pseudo-Hermitian Quantum Mechanics, 
arXiv:0810.5643 (2008).
%

\bibitem{DDTrev}
P. Dorey, C. Dunning and R. Tateo,
%
J. Phys. A: Math. Theor. 40, R205 (2007)
and

Pramana - J. Phys. 73, 217 (2009).


\bibitem{Dieudonne}
 J. Dieudonne,  
 Proc. Int. Symp.
            Lin. Spaces (Pergamon, Oxford,   1961), p. 115;

 J. P. Williams,
Proc. Amer.
            Math. Soc. 20,  121 (1969).

\bibitem{Chen}
C. M. Bender, S. F. Brandt, J.-H. Chen, and Q. Wang, Phys. Rev. D
71, 025014 (2005);
%
%
%
%

C. M. Bender, and P. D. Mannheim, Phys. Rev. Lett. 100, 110402
(2008).

\bibitem{timedep}
F. C. Figueira de Morisson, and A. Fring,
J. Phys. A: Math. Gen. 39,  9269 (2006);

%
%
A. Mostafazadeh,
Phys. Lett. B 650, 208 (2007);


M. Znojil, 
Phys. Rev. D 78, 085003  (2008).

\bibitem{alikg}
A. Mostafazadeh and F. Zamani,
Ann. Phys. (NY) 321, 2183  (2006) and 2210  (2006);

V. Jakubsk\'{y}, and J. Smejkal,
%
Czechosl. J. Phys. 56, 985 (2006).

\bibitem{Milos}
D. Krej\v{c}i\v{r}\'{\i}k, and M. Tater,
  J. Phys. A: Math. Theor. 41,  244013 (2008).

\bibitem{[24]}
B. Gutkin, and U. Smilansky, 
J. Phys. A: Math. Gen.  31, 6061 (2001);
%

R. Band, T. Shapira and U. Smilansky,
J. Phys. A.: Math. Gen. 39, 13999 (2006).
%


\bibitem{fragile}
M. Znojil, 
%
 J Math. Phys. 46, 062109 (2005).
%
%


\bibitem{cubic}
%
A. Mostafazadeh,  
J. Phys. A:
Math. Gen. 39,  10171 (2006).


\bibitem{Katka}
P. Exner and K. N\v{e}mcov\'{a}, 
J. Phys. A: Math. Gen. 36, 10173  (2003);
%
%

J. F. Brasche, and K. O\v{z}anov\'{a},
arXiv: math-ph/0511029;


K. O\v{z}anov\'{a},
J. Phys. A: Math. Gen. 39, 3071 (2006).

\bibitem{Dibwe}
D. P. Musumbu, F. G. Scholtz, and H. B. Geyer, J. Phys A: Math.
Theor. 40, F75 (2007);

C. Quesne,  J. Phys. A: Math. Theor. 41,  244022 (2008).

\end{thebibliography}
\end{document}